\documentclass[twocolumn,a4paper,prb]{revtex4-1}

\usepackage{color}

\usepackage{amsmath}
\usepackage{amsthm}
\usepackage{amssymb}
\usepackage{amsfonts}
\usepackage{layout}
\usepackage{graphicx}
\usepackage{dsfont}

\newtheorem*{2pFES}{Two-Point Correlator FES}
\newtheorem*{FES_approach}{FES Approach}
\newtheorem*{Direct_approach}{Direct Approach}

\newcommand{\ket}[1]{\left| #1 \right>} 
\newcommand{\bra}[1]{\left< #1 \right|} 
\newcommand{\braket}[2]{\left< #1 \vphantom{#2} \right|
 \left. #2 \vphantom{#1} \right>} 

\providecommand{\e}[1]{\ensuremath{\times 10^{#1}}}

\def\cA{{\cal{A}}}

\def\cD{{\cal{D}}}
\def\cG{{\cal{G}}}
\def\cO{{\cal{O}}}

\def\cG{{\cal{G}}}

\def\cH{{\cal{H}}}
\def\calZ{{\mathcal{Z}}}

\def\eye{\mathds{1}}

\def\Rbar{\overline{R}}

\def\Rdag{R^{\dagger}}

\def\conjR{\overline{R}}
\def\conjQ{\overline{Q}}

\def\hatpi{\hat{\pi}}
\def\hatphi{\hat{\phi}}
\def\hatpsi{\hat{\psi}}
\def\hatpsidag{\hat{\psi}^\dagger}
\def\psihdag{\hat{\psi}^\dagger}
\def\psih{\hat{\psi}}

\def\hata{{\hat{a}}}

\def\hatH{{\hat{H}}}
\def\hatcH{{\hat{\cal{H}}}}
\def\hatO{{\hat{O}}}
\def\hatT{{\hat{T}}}
\def\hatsigma{{\hat{\sigma}}}
\def\hatepsilon{{\hat{\epsilon}}}
\def\hatmu{{\hat{\mu}}}
\def\hatpsi{{\hat{\psi}}}
\def\hatpsibar{{\hat{\overline{\psi}}}}

\def\conjz{\overline{z}}
\def\conjw{\overline{w}}

\def\bh{\overline{h}}

\def\psidag{\psi^\dagger}

\def\vphan{\vphantom{\frac{1}{2}} }

\begin{document}
\title{Conformal Data from Finite Entanglement Scaling}

\author{Vid Stojevic$^{1}$, Jutho Haegeman$^{1}$, I. P. McCulloch$^{2}$ Luca Tagliacozzo$^{3}$, Frank Verstraete$^{1,4}$ }
\affiliation{
$^1$ Ghent University, Department of Physics and Astronomy, Krijgslaan 281- S9, B-9000 Ghent, Belgium \\
$^2$ School of Physical Sciences, The University of Queensland, Brisbane, QLD 4072, Australia \\
$^3$ ICFO The Institute of Photonics Sciences, Av. Carl Friedrich Gauss 3, E-08860 Castelldefels (Barcelona), Spain\\
$^4$ Vienna Center for Quantum Science, Universit\"at Wien, Boltzmanngasse 5, A-1090 Wien, Austria }

\begin{small}
\begin{abstract}

In this paper we apply the formalism of translation invariant (continuous) matrix product states in the thermodynamic limit to $(1+1)$ dimensional critical models. Finite bond dimension bounds the entanglement entropy and introduces an effective finite correlation length, so that the state is perturbed away from criticality. The assumption that the scaling hypothesis holds for this kind of perturbation is known in the literature as finite entanglement scaling. We provide further evidence for the validity of finite entanglement scaling and based on this formulate a scaling algorithm to estimate the central charge and critical exponents of the conformally invariant field theories describing the critical models under investigation. The algorithm is applied to three exemplary models; the cMPS version to the non-relativistic Lieb-Liniger model and the relativistic massless boson, and MPS version to the one-dimensional quantum Ising model at the critical point. Another new aspect to our approach is that we directly use the (c)MPS induced correlation length rather than the bond dimension as scaling parameter. This choice is motivated by several theoretical arguments as well as by the remarkable accuracy of our results.

\end{abstract}

\maketitle


\tableofcontents


\section{Introduction}

Matrix product states, both in their discrete \cite{Fannes:1992uq,verstraete2008matrix} and continuum \cite{PhysRevLett.104.190405,2012arXiv1211.3935H} variants, provide efficient descriptions of ground states of one-dimensional gapped systems. The reason for this is that the ground state of a local  gapped Hamiltonian in one dimension obeys an 'area law' \cite{1742-5468-2007-08-P08024, 1367-2630-12-2-025002},  a property that is built into the variational class. The area law, in any dimension, is the statement that the entanglement entropy of a large enough region scales not like the volume, but rather like the area of the boundary of that region\footnote{While it is expected that the area law holds for gapped systems in more than one spatial dimension, there exists no proof to this effect.}, which for (1+1) dimensional gapped systems means that the entropy of a large enough interval will saturate. 

At a critical point the gap goes to zero, and the low-energy behaviour of a one-dimensional system is described by a conformal field theory  (CFT) in $(1+1)$-dimensions. In this case the entanglement entropy of an interval increases proportionally to the logarithm of its length.  \cite{1994NuPhB.424..443H, PhysRevLett.90.227902, 1742-5468-2004-06-P06002} This implies that a matrix product state (MPS) or continuous matrix product state (cMPS) will not fully capture the behaviour of critical systems in the thermodynamic limit for any finite bond dimension. A different tensor network ansatz was constructed for critical systems by Vidal, the multi-scale entanglement renormalisation ansatz (MERA). \cite{2007PhRvL..99v0405V,2008PhRvL.101k0501V,2011arXiv1102.5524H} The structure of the MERA resembles the scale invariance present in critical ground states and supports the power-law decay of correlations. Indeed, a MERA description of a critical ground state allows to extract the critical exponents, both of local and nonlocal scaling operators and boundary scaling operators.\cite{2009PhRvA..79d0301P,2010PhRvB..82m2411E,2010PhRvB..82p1107E,2011arXiv1109.5334E}


Nevertheless, it was recently observed that the way in which a MPS approximation truncates the correlations in a critical ground state follows a universal scaling behaviour.\cite{2008PhRvB..78b4410T} This scaling was coined \emph{finite entanglement scaling}, as it is indeed the entanglement in the state which is bounded by the finite bond dimension $D$ of the (c)MPS approximation. As a typical (c)MPS has a finite correlation length, the (c)MPS approximation introduces a length scale which perturbs the CFT away from criticality. A scaling relation between this length scale and $D$ was obtained, which can be understood from interpreting $1/D$ as the distance from the critical point (which should be restored for $1/D=0$). An analytic expression for the corresponding critical exponent was first derived in Ref. ~\onlinecite{2009PhRvL.102y5701P} and then confirmed by independent calculations in Ref. ~\onlinecite{2012PhRvB..86g5117P} where also the crossover between the finite entanglement and finite size scaling in MPS with periodic boundary conditions was studied. Around the same time, one of the authors of this paper presented a direct approach to extract scaling exponents from MPS data. \cite{mcculloch_talks}  Since then, finite entanglement scaling has been used to find the phase diagram of spin models\cite{PhysRevB.87.235106,1751-8121-43-37-372001,1367-2630-13-2-023015} and to extract the CFT data from the edge theory of a fractional quantum Hall state.\cite{2013PhRvB..88o5314V} 

In this paper we provide further insight that helps clarify the validity of finite entanglement scaling, and enables us to develop an algorithm to estimate the central charge and critical exponents  of critical theories. In the next section we interpret FES using CFT ideas and formulate a \emph{scaling hypothesis}, which states how entanglement entropy and two-point correlation functions are expected to scale with bond dimension. The scaling hypothesis,  if valid, justifies the scaling algorithms for extracting the central charge and critical exponents of a CFT using (c)MPS presented in Section ~\ref{sec:D_scaling}.  These algorithms reduce to the method discovered by one of the authors \cite{mcculloch_talks} in a certain limit to be discussed. Unlike previous papers, we directly use the (c)MPS induced correlation length rather than the bond dimension as scaling parameter, and motivate the importance of this choice. Section \ref{section:examples} demonstrates these algorithms by applying them to three exemplary models:  1) the Lieb-Liniger model,  2) the massless relativistic boson in $(1+1)$ dimensions, and 3) the one-dimensional quantum Ising model at the critical point.  We apply our method both to CFT primary operators and also to a class of descendants. Remarkably for a (c)MPS based approach, for the massless relativistic boson our method is capable of estimating the exponents of vertex operators for arbitrary values of the real coefficient $\beta$\footnote{Up to restrictions imposed by the choice of UV regulator.} which parameterises a continuous infinity of distinct primary operators. The accuracy of the numerical results provides strong evidence for the scaling hypothesis. Another independent piece of evidence for finite entanglement scaling is provided by the observation that low-lying eigenvalues of the transfer matrix all scale in the same manner,  which implies that at large distances only a single independent scale is present in (c)MPS approximations of critical ground states. Section~\ref{sec:conclusions} presents our conclusions.  A brief review of (c)MPS is given in Appendix \ref{app:cMPS_review}. The field-field critical exponent calculation for the Lieb-Liniger model is presented in Appendix \ref{app:LL_field_field}, in order to illustrate the application of the algorithm presented in Section \ref{sec:D_scaling} in full detail. Finally, Appendix \ref{app:kappa} illustrates the importance of using the (c)MPS correlation length as scaling parameter for the accuracy of the results.


\section{Scaling hypothesis}
\label{sec:scalinghypothesis}
Several numerical methods for studying classical or quantum lattice systems are restricted to finite system sizes, due to the intrinsic finiteness of computer memory and computation time. Close to a critical point, the finite system size competes with the finite correlation length and the behaviour of thermodynamical quantities (\textit{e.g.} the order parameter or its susceptibility) can be modelled via scaling functions depending on the dimensionless quantity $L/\xi$. \cite{privman1990finite} Scaling at quantum critical points has been considered only recently. \cite{campostrini_finite-size_2014}

In a finite-size scaling approach (FSS), one would determine the scaling exponents of the different quantities by plotting the relevant quantities (e.g. the magnetisation) as a function of the dimensionless parameter and tuning the critical exponent such that the curves of these quantities extracted from different system sizes collapse.

For correlation functions depending on one or more spatial coordinates $x$, both $x/L$ and $x/\xi$ are dimensionless parameters and the scaling theory is more involved. However, exactly at the critical point ($\xi\to\infty$) of a $(1+1)$ dimensional system, the universal finite size effects can be obtained from the underlying conformal field theory (CFT).\cite{0305-4470-17-7-003} The crucial feature is that the predictions of a CFT are modified in a controlled way by mapping the theory originally defined on an infinite two dimensional plane to some other 2D geometry with a finite dimension such as, for example, an infinitely long cylinder with finite radius or an infinitely long strip with finite width.

For example, the footprint of a CFT, the power-law decay of correlation functions between a primary field of weight $(h, \bh)$ and itself on the infinite plane,
\begin{align}
& \bra{0} \hatO_A (z_1) \hatO_B (z_2 ) \ket{0}  =:  \cG_{\hatO}(z_{12})  \\ \nonumber 
& = \frac{ 1  }{(z_{12} )^{2h} (\conjz_{12} )^{2 \bh} } \ ,
\end{align}
is modified to 
\begin{align}
 \label{eq:cyl_correlator}
& \cG_{(L) \hatO} (w_{12})   =  \left( \frac{2 \pi}{L} \right)^{2(h + \bh)}  \times\\ \nonumber
& \left(  2 \sinh \left(\frac{ \pi (w_{12} )  }{L} \right)    \right)^{-2h}  \
 \left(  2 \sinh \left(\frac{ \pi (\conjw_{12} )  }{L} \right)    \right)^{-2\bh} \ 
\end{align}
on the cylinder, where $z_{12} := z_1 - z_2$ and similarly for $w_{12}$.\footnote{ Here $z : = x_0 + i x_1$, where $x_0$ and $x_1$ are the euclidean space and time coordinates, and the infinite plane coordinates $z$ are related to coordinates on the cylinder as $z = e^{\frac{2 \pi w}{L}}$. The time direction on the cylinder corresponds to the radial direction on the infinite plane, with the origin mapping to the infinite past, while the angular direction on the infinite plane corresponds to moving along the finite direction of the cylinder. The current discussion contains a lot of standard vocabulary used when working with (1+1) CFTs. A reader unfamiliar with the subject can consult, for example, the standard reference \cite{citeulike:1280772}.}

Low energy properties  at a conformally invariant critical point are equally well described by considering a classical two dimensional system or an equivalent one dimensional quantum system. For this reason, finite size effects are also observed in genuinely quantum properties such as in the scaling of the entanglement entropy. The entanglement entropy of an interval of length $x$ belonging to a chain of length $L$ with periodic boundary conditions is indeed described by: 
\begin{align}
\label{eq:cyl_entropy}
S_{(L)} = \frac{c}{3} \log \left( \frac{L}{\pi a} \sin \left(\frac{\pi x}{L} \right) \right) + k \ .
\end{align}
In the limit $L \rightarrow \infty$, or for small $x\ll L$, one recovers the well-known thermodynamic limit expression  \cite{1994NuPhB.424..443H, PhysRevLett.90.227902, 1742-5468-2004-06-P06002}:
\begin{align}
\label{eq:entropy_vs_logmu}
S =   \frac{c}{3} \log (x) + k \ .
\end{align}
 
 The crucial observation is that the finite size effects in both expressions (\ref{eq:cyl_correlator}, \ref{eq:cyl_entropy}) enter via a function that depends on distance in units of $L$, i.e. via $x/L$ in the case of entropy and $w_{12} /L $ in the case of the two-point correlator. Similar expressions exist when not the spatial size but the temporal size of the system is finite (\textit{i.e.} finite temperature).

A (c)MPS based FSS approach for one-dimensional critical theories would make use of the fact that finite size introduces a gap in the CFT so that its ground state can be well captured by the variational manifold, provided that the bond dimension grows sufficiently rapidly with the system size.\cite{2012PhRvB..86g5117P} A natural approach for calculating the central charge $c$ or the critical exponents $\Delta : = h + \overline{h}$ using (c)MPS is as follows. First pick a range of circles on which the spatial direction of the CFT is "compactified", and for each calculate the (c)MPS ground state at large enough bond dimension to adequately capture the exact ground state. Next pick a scale $s<1$, and for each circle calculate the entropy of an interval of length $x = s L$ numerically using (c)MPS. It is obvious that for \emph{any} choice of $s<1$ one can obtain an estimate for $c$ from the scaling of $S_{(L)}$ vs. $\log(L)$.  Similarly, critical exponents can be estimated from from the scaling of $\log( \cG_{(L)\hatO})$ vs. $\log(L)$. Since both $S_{(L)}$ and $\cG_{(L)\hatO}$  are calculated from  (c)MPS data, for numerical reasons some values of $s$ may be preferred, and we can scan over $s$ in order to obtain the best numerical fit.

Now let us imagine that a length scale $\mu$ is introduced via some other mechanism, either with or without a geometric origin. It is obvious that the scaling approach described above can be applied 'as is' regardless of the manner in which this scale is introduced, as long as the effect on entanglement entropy and two-point correlator expressions is through a \emph{scaling function} of $(x/\mu)$ such that:
\begin{align}
\label{eq:entropy_scaling_general}
S_{(\mu)}(x, D) \propto \frac{c}{3} \log \left( \frac{\mu }{\pi a} f \left( \frac{x}{\mu } \right) \right)
\end{align}
and 
\begin{align}
\label{eq:2pcorr_scaling_general}
\cG_{(\mu) \hatO} (z_{12}) \propto \left( \frac{1}{\mu} \right)^{2(h + \bh)} g \left( \frac{x}{\mu } \right) \ .
\end{align}
The precise form of the functions $f$ and $g$ is immaterial; the central charge can be calculated from the scaling of $S_{(\mu)}$ versus $\log(\mu)$, and the critical exponents from $\log( \cG_{(\mu)\hatO})$ versus $\log(\mu)$. Equations (\ref{eq:entropy_scaling_general}) and (\ref{eq:2pcorr_scaling_general}) constitute our \emph{scaling hypothesis}.

Recently (c)MPS methods have been developed that enable the study of physical systems directly in the thermodynamic limit \cite{PhysRevLett.98.070201, PhysRevLett.107.070601, 2012arXiv1207.0691M, 2012arXiv1211.3935H, 2013PhRvB..88g5133H, 2008arXiv0804.2509M}, using a translation invariant Ansatz; for MPS,
\begin{align*}
& \ket{ \Psi [ A ] }= \\ \nonumber & \sum_{i_1=1}^d \sum_{i_2=1}^d \cdots \sum_{i_N=1}^d v_L^\dagger A^{i_1}_1 A^{i_2}_2 \cdots A^{i_N}_N  v_R \ket{ i_1, i_2, \cdots, i_N } 
\end{align*}
($N\rightarrow \infty$) with $A$ position independent, and for cMPS
\begin{align}
\label{eq:ucMPS2}
& \ket{ \Psi [ Q(x), R_{\alpha}(x) ] } = \\ \nonumber 
&  v_L^\dagger \mathcal{P} \mathrm{exp} \left[  \int_{- \frac{L}{2}}^{\frac{L}{2}} dx   \ \left( Q(x) \otimes \eye + \sum_\alpha R_{\alpha} \otimes \hatpsidag_\alpha (x)  \right)  \right]  v_R  \ket{ \Omega}   \ ,
\end{align}
($L\rightarrow \infty$) with $R$ and $Q$ position independent. The long distance behaviour of correlation functions with respect to a (c)MPS is governed by the second largest eigenvalue $\lambda_2$ of the transfer matrix $T$ [defined in Eq.~\eqref{eq:transfer_matrix} for cMPS and in Eq.~\eqref{eq:MPS_transfer_matrix} for MPS]; the largest eigenvalue is required to be zero in order to ensure correct normalisation. The finite bond dimension $D$ thus introduces a finite correlation length:
\begin{align}
\label{eq:corr_length_def}
\mu_2 (D) = - \frac{1}{\lambda_2(D)} \ ,
\end{align}
which perturbs the state away from the critical point.  It was demonstrated in  Ref.~\onlinecite{2008PhRvB..78b4410T} that the effective correlation length asymptotically scales as $\mu_2(D)\sim D^{\kappa}$, where $\kappa$ is a constant that depends only on the universality class.  As the bond dimension bounds the maximal entanglement in the state,  this kind of scaling is also referred to as \emph{finite entanglement scaling} (FES). Once the exponent $\kappa$ has been determined Ref. ~\onlinecite{2008PhRvB..78b4410T} outlines an approach, different to the one presented in this paper, for extracting  critical exponents by performing a scaling analysis directly with respect to $D$.  While the precise manner in which perturbation due to finite bond dimension affects the CFT is not properly understood, these results constitute evidence that the scaling hypothesis (\ref{eq:entropy_scaling_general}, \ref{eq:2pcorr_scaling_general}) holds for FES. 

 Assuming the validity of this scaling relation, the exponent $\kappa$ was later determined in function of the central charge $c$ of the CFT as \cite{2009PhRvL.102y5701P}:
\begin{align}
\label{eq:kappa_def}
\kappa = \frac{6}{c \left( \sqrt{\frac{12}{c}} + 1 \right)} \ .
\end{align}
In this paper we provide further evidence in favour of the FES hypothesis  by observing the higher eigenvalues of the transfer matrix, which also induce a length scale $\mu_I(D) = -1/\Re(\lambda_I(D))$ for $I>2$. Our numerics reveal that ratios of the real parts of the low-lying eigenvalues of the transfer matrix $T$ are roughly constant. This is demonstrated in Figure~\ref{fig:Ising_combined} for the quantum Ising model at the critical point. The fact that all the eigenvalues of the transfer matrix obey the same scaling is a further hint that equations like the ones in (\ref{eq:cyl_correlator}, \ref{eq:cyl_entropy}), which ultimately are consequences of the presence of a single scale, could also describe FES. The (one-parameter) scaling hypothesis  (\ref{eq:entropy_scaling_general}, \ref{eq:2pcorr_scaling_general}) would be violated if  different eigenvalues of the transfer matrix would scale with different powers of $D$, thus producing several independent relevant infrared length scales. 

In order to attempt to understand this observation, let us imagine that the finite bond dimension induced scale has some geometric origin or interpretation. An initial tempting guess, which is ultimately too simplistic, might be to postulate that the (c)MPS transfer matrix represents the contraction of a section of a 2D tensor network encoding  the partition function of a related classical model on an infinite strip (since the (c)MPS describes an infinite chain  with finite width). This would mean that the (c)MPS transfer matrix is equivalent to  the transfer matrix of the classical model  along the infinite direction on the strip. For this geometry the ratios of the eigenvalues of the transfer matrix are known and independent of the scale, i.e. the width of the strip.\cite{cardy_operator_1986,cardy_effect_1986} It is however not clear that the origin of the finite entanglement scale really is geometric, and our numerical results for the ratios of  the eigenvalues of the transfer matrix do not reproduce  the ones expected from the corresponding CFTs on the strip. Nevertheless, the fact that the ratios converge to a well defined scale independent value is another piece of evidence that there should be  a CFT interpretation of FES.\footnote{A more sophisticated CFT interpretation of FES would be in terms of a strip with a line of impurities bisecting it along the infinite dimension, where the impurity line is related to the gluing of the (c)MPS with its complex conjugate. At present we do not have a good enough understanding of the effect of such an impurity line to say whether or not this proposal is correct.}

\begin{figure*}[!htb]
\centering
\includegraphics[width=16cm]{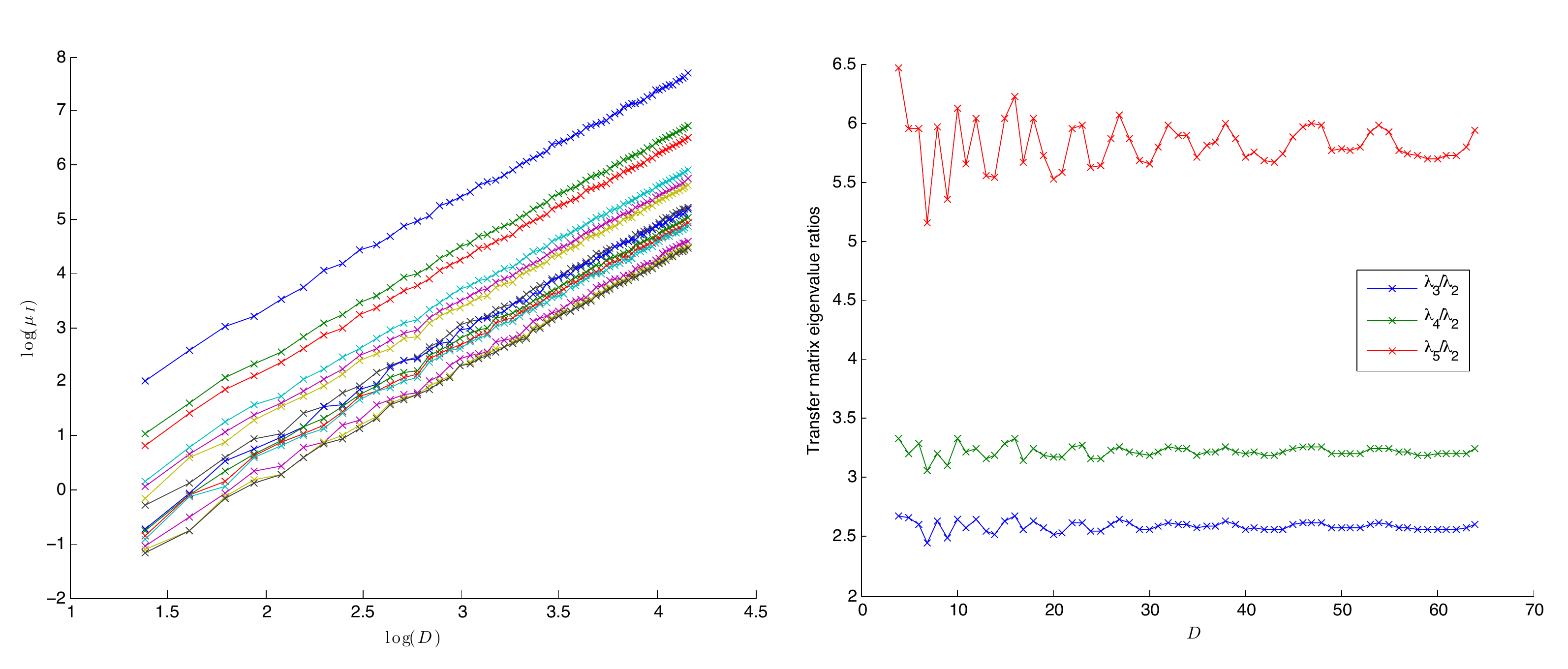}
\caption{\footnotesize{(Colour online). The left plot depicts the linear relation between $\log(\mu_I)$ and $\log(D)$ for low lying eigenvalues of the MPS transfer matrix of the critical quantum Ising model (the largest fifteen non-zero eigenvalues are displayed). The right plot shows that the ratios of the eigenvalues approximately converge to fixed values  with increasing bond dimension $D$. The low lying eigenvalues of the cMPS transfer matrices for the two field theory models studied in this paper, the Lieb-Liniger and the massless relativistic boson, display the same general behaviour. The eigenvalue ratios converge to different values for the three models.  A reminder of the notation: the eigenvalues of the transfer matrix are denoted by $\lambda_I$, with $\lambda_1 = 0$, $\lambda_2$ the  largest non-zero eigenvalue, and $\mu_I = -1/\lambda_I$. } }  \label{fig:Ising_combined}
\end{figure*}

This paper presents a scaling algorithm based on the FES hypothesis (\ref{eq:entropy_scaling_general}, \ref{eq:2pcorr_scaling_general}). Unlike previous papers that use $D$ or $D^\kappa$  as scaling parameter, our approach directly uses the (c)MPS induced correlation length $\mu_2(D)$ as scaling parameter. There are several benefits to this approach. As $\mu_2(D)$ has the dimension of a length scale, it is the most natural parameter to be used in the scaling relations (\ref{eq:entropy_scaling_general}, \ref{eq:2pcorr_scaling_general}). Secondly, even when the parameters of the Hamiltonian are slightly different from its critical point (\textit{e.g.} because the precise location is not exactly known), we can still argue that $\mu_2(D)$ is the only relevant length scale in the system. While the $D$-limited length scale $D^\kappa$ would compete with the physical correlation length $\xi$ resulting in a two-scale problem, we anticipate that the observed correlation length $\mu_2(D)$ automatically interpolates between these two length scales in such a way that the scaling relations (\ref{eq:entropy_scaling_general}, \ref{eq:2pcorr_scaling_general}) continue to hold. Another significant problem with scaling using the bond dimension is that converging to an optimum ground state is computationally very expensive. Often one is much better increasing $D$, even by a small amount, and doing a few iterations of TDVP (or iDMRG) rather than doing many iterations to reach the true optimum for smaller $D$. This provides another significant advantage to using the correlation length in practical calculations. Finally, as is shown in Appendix \ref{app:kappa}, the scaling approach based on $\mu_2(D)$ as scaling parameter produces more accurate results for the critical exponents and central charge.


\section{Recipe for Finite Entanglement Scaling}
\label{sec:D_scaling}
In this section we describe a finite entanglement scaling (FES) method for estimating critical exponents and the central charge of a conformally invariant theory.

\subsection{Critical Exponents}

 Two-point correlation functions in critical theories obey power-law decay at large distances, in contrast to the exponential falloff that occurs for gapped models. That is, in a CFT, the two-point correlation function of a primary operator $\hatO$ with itself behaves as
\begin{align}
\label{eq:power_law}
\cG_{\hatO}(x) = \bra{0} \hatO^{\dagger} (0) \hatO (x)  \ket{0}  \propto x^{-2 \Delta_{\hatO}}  \ \ \ , \ \ \ x \gg 0 \ ,
\end{align}
where $\Delta_{\hatO}$ is the critical exponent corresponding to $\hatO$.  We will not be considering correlation functions between different operators in this paper.
 
The cMPS approximation of the CFT ground state at any finite bond dimension $D$ generates a gap, and the approximation of the two-point correlation function,
\begin{align}
\label{eq:power_law_cmps}
G_{\hatO}(x) : = (l | O^{\dagger} [ R, \conjR, Q, \conjQ ]  e^{T x}  O [ R, \conjR, Q, \conjQ] | r ) \ ,
\end{align}
 reproduces the power-law decay up to some distance generally shorter then, or at best of the order of the correlation length (as defined in Eq.~\eqref{eq:corr_length_def}), and decays exponentially beyond that (see Figure \ref{fig:LL_psi_disconn_correlator} in Appendix~\ref{app:LL_field_field}).\footnote{In this section we will use the conventions and language appropriate for continuous systems. Two-point correlators for lattice systems  can clearly only be calculated with the distance between operator insertions a multiple of the lattice spacing. For the purposes of the scaling calculations we interpolate in order to obtain correlator values at arbitrary points. The dependence of the critical exponent estimates on the type of interpolation used (we have compared linear and spline interpolations) is negligible at the large distances at which the FES estimates are obtained. The reason for this is that a scaling calculation, together with the interpolation subroutine, is concerned with the logarithm of the correlator as a function of the logarithm of distance, so at scales much larger than the lattice spacing neighbouring points are very near to each other. Statements made in this section are therefore equally valid for lattice systems once the interpolation step is performed\label{ftn:interpolation}}

The observation central to our algorithm for approximating critical exponents is a consequence of the scaling hypothesis (\ref{eq:entropy_scaling_general}, \ref{eq:2pcorr_scaling_general}):
 \begin{2pFES}
At all scales $s$ large enough to eliminate short distance artefacts,  $\log(G_{\hatO}(s \mu_2(D))$ scales linearly with respect to $\log( \mu_2(D))$ with the constant of proportionality given by $-2 \Delta_\hatO$.
 \end{2pFES}
 
Using this property, critical exponents can be estimated as follows:
\begin{FES_approach}
Using (c)MPS approximations for the critical ground state for a range of bond dimensions,  estimates for $\Delta_{\hatO}$ at different scales $s$ are given by the slopes obtained from the linear interpolation of  $\log(G_{\hatO}(s \mu_2(D)) )$ vs. $\log (s \mu_2(D))$. We scan over $s$ such that $s_0 < s < \infty$, and $s_0$  is large enough to wash out any  short distance/cutoff effects.  The final result for the exponent is  obtained from the interpolation of $\log(G_{\hatO}(s \mu_2(D)) )$ vs. $\log (s \mu_2(D))$ at the scale $s$ at which the confidence interval for the the slope is minimal. The error estimate for the exponent is given by the confidence interval.  
 \end{FES_approach}
The confidence intervals for the slopes depend on the choice of the confidence level; in this paper we will calculate error estimates for the slopes using both $95 \%$ and $99.73\%$ confidence levels. It is not obvious that the scan over $s$ improves the accuracy of the estimates, over simply choosing some particular value, e.g. $s=1$ or considering the limit $s\rightarrow \infty$, but it turns out that this is numerically a worthwhile step.

Using the eigenvalue decomposition of $T$ and writing the distance $x$ in units of $\mu_2$
 \begin{align}
\label{eq:s_def2}
 x =  s \mu_2(D) \ ,
 \end{align}
 (\ref{eq:power_law_cmps}) can be re-expressed as:
\begin{align}
\label{eq:correlator_expanded}
G_{\hatO}(s \mu_2(D)) ) = | ( l | O |r ) |^2 +  \sum_{I=2}^{D^2} ( l | O^\dagger |r_I ) e^{-s \frac{\lambda_I}{\lambda_2} } (l_I | O |r)  \ .
\end{align}
Here $(l_I | $ and $|r_I)$ are the left and right eigenvectors corresponding to the eigenvalue $\lambda_I$; $( l_1 | \equiv (l |$ is the zero-eigenvalue eigenvector. We have suppressed the $D$ dependence of eigenvectors and eigenvalues on the right hand side. At $s=\infty$ only the dominant contribution to $G_{\hatO}(s \mu_2(D)) ) $ survives. Let us suppose that the first non zero contribution is for $I=a$, then in the limit of large $s$ scaling $ \log( \exp(-s \frac{\lambda_a}{\lambda_2}) ( l | O^\dagger |r_a ) (l_a | O |r) ) $ vs. $\log( \mu_2(D) )$ provides an estimate for the critical exponent. If the first non-zero contribution is for $a=2$, the prefactor is constant. If on the other hand it occurs at some $a >2$, it is still roughly constant, since the ratios of the eigenvalues converge (see Figure  \ref{fig:Ising_combined}). However, since the low lying eigenvalues all scale in the same way, the FES approach described above is also valid with any low lying $\mu_a$ replacing $\mu_2$. It follows that dropping the prefactor in front of the dominant contribution is ok, i.e. that simply scaling  $\log (  ( l | O^\dagger |r_a ) (l_a | O |r) )  )$ vs. $\log( \mu_2(D) )$  should provide an estimate for the exponent.  This indeed turns out to be the case, as was observed by one of the authors of this paper.\cite{mcculloch_talks}   However, for nearly all the calculations performed in this paper, estimates obtained at $s$ of the order of one, that contain all contributions from an arbitrarily large number of eigenvectors of the transfer matrix, are superior to the fits at $s= \infty$.


 The remaining problem at this stage is how to determine $s_0$, or at least an upper bound for it. To address this problem, let us first consider estimates for the exponents obtained directly from the (c)MPS  approximation to the correlator at one particular bond dimension:
\begin{Direct_approach} At one particular bond dimension pick a distance $x_I < \mu_2$ at which the algebraic decay is well captured by the (c)MPS approximation, but which is still large enough to wash out any short distance/cutoff effects. Estimate $\Delta_{\hatO}$ from the slope of $\log(G_{\hatO}(x) )  )$ vs. $\log(x)$ at $x_I$. 
\end{Direct_approach}

The relation between the Direct and FES approaches is demonstrated in Figure \ref{fig:LL_field_plot3d} for the ($\hatpsi-\hatpsidag$) correlator in the Lieb-Liniger model at  $g_{\mathrm{eff}} = 1.348..$ . It is clear that an algorithm based on  the Direct Approach alone is beset by serious obstacles, the most serious being  that no general method exists to determine the window for $x_I$ inside which estimates are accurate. In addition, estimating the error in the estimates is not as straightforward as in the FES scalings.   One could attempt to overcome these problems by working with a set of bond dimensions  and choose the critical exponent estimate corresponding to the scale at which the spread  in estimates is minimal.   Unfortunately it turns out that the minimal spread often occurs in regions where short distances  effects are important thus in general missing the true value of the exponent.

\begin{figure}[!htb]
\centering
\includegraphics[width=8cm]{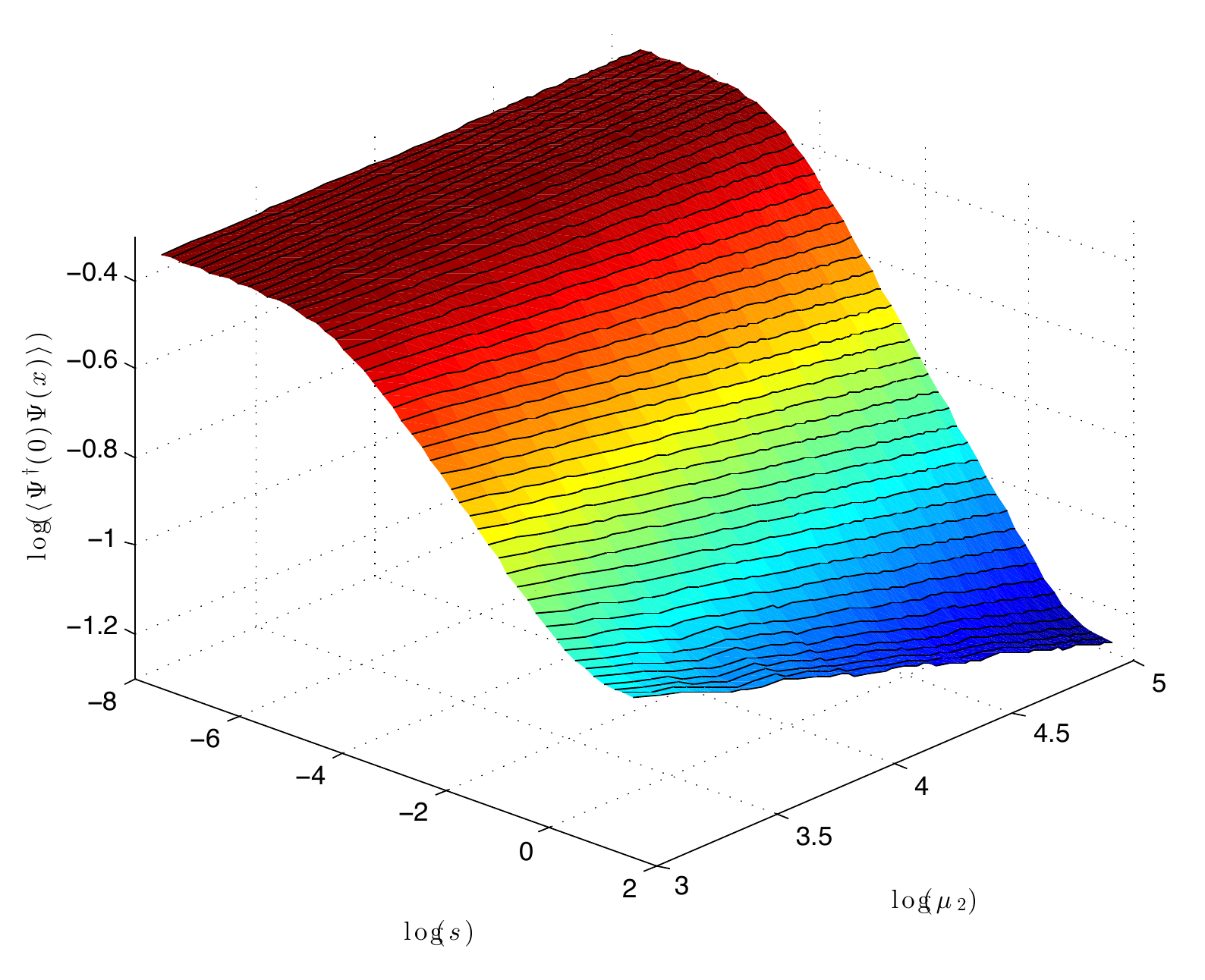}
\caption{\footnotesize{ (Colour online). The logarithm of the field-field ($\hatpsi-\hatpsidag$) correlator of the Lieb-Liniger model as a function of the logarithm of the scale $s$ ($s$ being the distance between the $\hatpsi$ insertions in units of the correlation length $\mu_2(D)$) and  $\log( \mu_2 (D))$.  The plot displays results for $0.1 \leq s \leq 4$ and $14 \leq D \leq 64$. Plotting the logarithm of the correlator  as a function of $\log (\mu_2(D))$ (which is a monotonic function of bond dimension), rather than directly as a function of $D$, is more revealing for our purposes since a slice in the $\log( \mu_2 (D)) = const $ plane provides the relevant information for the \emph{Direct Approach} (taking slices of the above surface at  $\log(\mu_2(D))$ corresponding to $D=16,24,32,48,64$, and rescaling the units of distance appropriately at each bond dimension, yields the curves displayed in Figure \ref{fig:LL_psi_disconn_correlator}). A slice in the $\log(s) = \mathrm{const} $ plane on the other hand is what is relevant when applying the \emph{FES Approach}. A slice at $s=0.66$,  which turns out to be the scale at which the optimum estimate for the $\hatpsi-\hatpsidag$ critical exponent is obtained, is given in Figure \ref{fig:LL_field_logcorr_vs_logpos_with_fit}. The above plot gives a rough visual impression of the fact that the Direct Approach is only accurate in the linear region below $s=1$, while the FES approach is also accurate for $s > 1$. Both approaches break down at short distances due to non-universal/cutoff effects.   } } \label{fig:LL_field_plot3d}
\end{figure}

We do proceed by working with a range of bond dimensions $\cD$, and apply the Direct Approach at each of these, but use this simply in order to get an upper bound on $s_0$ for the  FES Approach.  Above $s=1$ the critical exponent estimates in the Direct Approach will be completely off, as the algebraic falloff is no longer captured by the (c)MPS approximation to the correlator. The approximations will become more accurate at some distance below the correlation length, and will again become unreliable at short distances. Since the FES Approach remains accurate for $s>1$, an upper bound for $s_0$ in the FES approach is given by the maximum scale below $s=1$ at which the FES and Direct Approach method results intersect.  A region in which the two approaches agree is expected to exist in general, since in some region below the correlation length $G_{\hatO}(x)$ will have converged to good accuracy for all bond dimensions used in the FES Approach (see e.g. the plot in Figure \ref{fig:LL_psi_disconn_correlator}). There are exceptions to this, that is, cases when no clear intersection exists. This can occur, for example, when for all bond dimensions in $\cD$ the Direct Approach estimate only approaches the true value for all bond dimensions in the range, but never reaches it, and then deviates wildly at very small $s$. In such cases we simply have to restrict our FES scan from $s=1$ to $s=\infty$, i.e. we work with $s_0 = 1$, which generally still brings about a large increase in accuracy over the estimate at $s=\infty$. For the field theory examples studied in this paper we always see a clear intersection, but this is not the case for all the operators in the Ising model example (see Table \ref{tab:ising_results}).

We now have all the ingredients to for a robust algorithm to calculate critical exponents:
\begin{enumerate}
\item{For a  set of bond dimensions $\cD$, apply the FES Approach scanning all scales from zero to infinity\footnote{In practice this means scanning  from sufficiently close to zero that short distance effects are obvious, to far enough beyond the correlation length that only the dominant eigenvector contribution remains.}, and store the critical exponent estimates at all scales. Having chosen an appropriate confidence level, the error bars are determined by the confidence interval for the slope. }
\item{For all all the bond dimensions in $\cD$ apply the Direct Approach, scanning over all distances from zero to infinity. At each bond dimension store the estimates for all $s$.}
\item{   Take $s_0$ to be the maximum scale at which the estimates from 1) and 2) agree. The final estimate of the critical exponent is given by the FES estimate with the smallest confidence interval for the slope in the range $s_0 \leq s < \infty$. }
\end{enumerate}

\subsection{Central Charge}

For a (c)MPS the density matrix corresponding to an interval of length $x$ is given by  the $D^2 \times D^2$ matrix:
\begin{align}
\rho = (l^T)^{\frac{1}{2}} \otimes  (r^T)^{\frac{1}{2}} (\widetilde{\exp (T x)})^{\frac{1}{2}}  \ ,
\end{align}
where
\begin{align}
\widetilde{\exp (T x)}_{ijkl} := \exp (T x)_{ikjl}  \ .
\end{align}
Here $l$ and $r$ are the left and right  zero-eigenvalue eigenvectors of the transfer matrix  $T$ reshaped into $D\times D$ matrices (see Appendix \ref{app:cMPS_review} for more details). The corresponding entanglement entropy is given by
\begin{align}
S =  - \mathrm{tr} ( \rho \log (\rho ))  = -  \sum_i \lambda_i^2 \log ( \lambda_i^2) \ .
\end{align}
where  $\lambda_i$ are the Schmidt coefficients corresponding to $\rho$. Following the discussion in the context of (\ref{eq:entropy_scaling_general}), after choosing a scale $s$, the central charge can be estimated from the scaling of $S(D)$ of an interval $x(D) = s \mu_2 (D)$ vs. $\log( \mu_2 (D) )$. The error estimates are again given by the confidence interval for the slope, and depend on the choice of the confidence level. Since $S$ is obtained numerically from the (c)MPS data, a different estimate for $c$ is in general obtained at each scale  $s$. 

For the examples studied in this paper we observe, by comparing to exact results, that the linearity of the scalings based on the interval entanglement entropy improves down to some scale $s_{\mathrm{opt}} < 1$, below which it becomes inaccurate due to short-distance/cutoff effects. When determining critical exponents we encountered a similar problem of having to determine an optimum scale, and made use of estimates obtained directly from the (c)MPS approximation of the two-point correlation function at some fixed bond dimension in order to give an upper bound for the optimum scale and ensures that we do not pick a scale that is too small. An analogous approach is also possible for the calculation of the central charge.  Unfortunately the computational cost of calculating the entanglement entropy of a finite interval is $\cO (D^6)$,  so scanning over $s$ becomes a lot more expensive than for the critical exponent calculations, where the computational cost is only $\cO (D^3)$. We have not found it feasible to implement such an algorithm for the models considered in this paper. In addition, unlike for critical exponent estimates where the increase in accuracy over $s = \infty$ is already significant for $s$ close to $1$, the analogous gain in accuracy for estimates of central charge turns out to be very poor (in particular this means that scaling using intervals at the value $s=1$, where we need not worry about short distance effects, gives virtually no improvement in accuracy).

For these reasons, instead of working with the entanglement entropy of an interval, as given by equation (\ref{eq:entropy_vs_logmu}), we will  consider a bi-partition of a finite system and the entanglement entropy of the half-system $\cA$. In the limit of growing the length $x_{\cA}$ of $\cA$ to infinity, the entropy of the half-system now grows as:
\begin{align}
\label{eq:entropy_vs_logmu_open}
S =   \frac{c}{6} \log (x_{\cA}) + k \ .
\end{align} 
The simplest approach to calculating the central charge is indeed by using the half-infinite line entanglement entropy rather than entropy of an interval, since the density matrix of a half-infinite line (chain) in the (c)MPS approximation is only $D \times D$ dimensional:
\begin{align}
\rho  = (l^T)^{\frac{1}{2}} r^{\frac{1}{2}} \ .
\end{align}
One can easily check that the contributions to the interval entanglement entropy due to non-zero eigenvalue eigenvectors of $T$ vanish as the interval is taken to infinity, and that the interval and half-infinite line estimates for $c$ become equal in the limit $s \rightarrow \infty$.


We have also examined the possibility of exploiting the conjectured relation between $D$ and $\mu$  \cite{PhysRevLett.107.070601, 2008PhRvB..78b4410T, 2008PhRvA..78c2329C, PhysRevLett.56.742}, namely that:
\begin{align}
\label{eq:Dscaling}
\mu_2 \propto D^{\kappa} \ ,
\end{align}
 with $\kappa$ analytically determined as function of $c$ in Eq.~\eqref{eq:kappa_def}.
Using this relation the central charge can be estimated from the slope of $\log(\mu_2 (D))$ vs. $\log(D)$. Another estimate for $c$ can be obtained by combining the half-infinite entropy with (\ref{eq:Dscaling}, \ref{eq:kappa_def}), so:
 \begin{align}
 \label{eq:entropy_vs_D}
 S = \frac{1}{\sqrt{\frac{12}{c}} + 1} \log (D)  + k\ ,
 \end{align}
 and  $c$ can be estimated also from the scaling of $S(D)$ vs. $\log(D)$. Alternatively, we can keep $\kappa$ as a free parameter and work simply with:
  \begin{align}
 \label{eq:entropy_vs_D2}
S =   \frac{\kappa c}{6} \log (D) + k \ .
\end{align} 
 That is, we still obtain $c$ from the scaling of $S(D)$ vs. $\log(D)$, but use the value for $\kappa$ obtained from the scaling of $\log(\mu_2 (D))$ vs. $\log(D)$ instead of  using (\ref{eq:kappa_def}). The interval entanglement entropy grows twice as quickly with  $\log (D)$ compared to expressions (\ref{eq:entropy_vs_D}, \ref{eq:entropy_vs_D2}).


In this paper we obtain estimates for $c$ for the three aforementioned  models using the half-infinite line entropies. For the two field theories we also obtain estimates based on interval entropies at $s=0.1$, in order to demonstrate that an increase in accuracy is obtained by going to finite $s$, albeit a modest one. We observe significant deviations form the predicted value for $\kappa$  (\ref{eq:kappa_def}) for all three models, and estimates for $c$ that depend on this relation turn out to be inaccurate. Central charge estimates obtained from scalings with respect to $\mu_2(D)$ are presented in the next section, and those obtained from scalings with respect to  $D$ directly are presented in Appendix \ref{app:kappa}.

\section{Exemplary Models}
\label{section:examples}
 
 In this section we consider three exemplary critical models in order to demonstrate the FES approach to calculating the central charge and critical exponent that was described in the previous section. A cMPS version of the algorithm is applied to the Lieb-Liniger model \cite{PhysRev.130.1605, PhysRev.130.1616}, which describes an interacting non-relativistic one dimensional Bose gas, and also to the relativistic massless boson in $(1+1)$ dimensions. The MPS version is applied to the one-dimensional quantum Ising model at the critical point. The scaling calculations for all three models are performed using all bond dimensions in the range $ 32  \leq D \leq 64$. The (c)MPS approximations of the ground state are obtained using the time dependent variational principle \cite{2013PhRvB..88g5133H} combined with a conjugate gradient method\cite{2013PhRvD..88h5030M}; for the MPS case an equally efficient option is to use the infinite-size variant of the standard DMRG algorithm (iDMRG). \cite{2008arXiv0804.2509M}

 \subsection{Lieb-Liniger Model}
 \label{subsection:LL}

The Lieb-Liniger model describes bosons on a line interacting via a contact potential. The Hamiltonian is given by:
\begin{align}
\hatH = \int_{-\infty}^{\infty} dx \left[ \frac{d}{dx} \hatpsi^\dagger \frac{d}{dx} \hatpsi + v  \ \hatpsidag \hatpsi + g  \ \hatpsidag \hatpsidag \hatpsi \hatpsi \right]  \ , 
\label{eq:LL_hamiltonian}
\end{align}
and the theory is critical for the whole range of parameters $g >0$, $v < 0$. The effective space of vacua is not two-dimensional, as the only relevant parameter is the effective interaction strength $g_{\mathrm{eff}} := g / \rho^2$, where $\rho$ is the particle density, and $g_{\mathrm{eff}}$ can be adjusted by either changing the chemical potential $v$ or the interaction strength $g$.  The central charge of the Lieb-Liniger model is known to be $c=1$.


In this section we consider the ground state of the Hamiltonian (\ref{eq:LL_hamiltonian}) with $v=1,  g=1$, which corresponds to $g_{\mathrm{eff}} = 1.348...$. 

We observe that the low lying eigenvalues of the transfer matrix of the Lieb-Liniger model all scale in the same manner (see discussion at the beginning of Section \ref{sec:D_scaling}). The situation is very similar to that depicted for the quantum Ising model in Figure \ref{fig:Ising_combined}, except that the ratios converge to different values. Estimates for $\kappa$, as obtained from the scalings of  $\log (\mu_I)$ vs. $\log(D)$ (see Eq. (\ref{eq:Dscaling})),  underestimate the predicted value (\ref{eq:kappa_def}) for all $I$. The value obtained from the scaling of $\log (\mu_2)$ vs. $\log(D)$ is given in Table \ref{tab:kappa_results} in Appendix \ref{app:kappa}.

We also obtain estimates for $c$ using scalings of $S$ vs. $\log(\mu_2)$, using both the entanglement entropy of the half-infinite line, and also of finite intervals of length $0.1 \mu_2(D)$ (i.e. at $s=0.1$).  We have not implemented  a robust method for obtaining a lower bound for $s$, due to the high resources necessary for such a computation and the very modest gain in accuracy (see discussion in Section \ref{sec:D_scaling}). That is, we do not give any demonstration that the value $s=0.1$ is large enough so that cutoff effect are not present, independent of the fact that the know exact value $c=1$  is reproduced. The results for $s=0.1$ demonstrate at least that the accuracy \emph{can} be improved over the scaling at $s=\infty$.  There is an improvement already when picking the "safe" value $s=1$, but this improvement turns out to be so small that  it is negligible, at least for the range of bond dimensions we are using.

Central charge estimates obtained from using half-infinite line entropies are summarised in Table \ref{tab:c_half_line_results}, and from entropies of intervals of length  $0.1 \mu_2(D)$ in Table \ref{tab:c_LL_results_interval}.  Critical exponent estimates have been obtained for a number of Lieb-Liniger operators and are listed in Table \ref{tab:LL_results} - various details pertaining to the particular operators are presented in the remainder of this subsection.

 As a guiding example for the method, the field-field exponent calculation is spelled out in full detail in Appendix \ref{app:LL_field_field}.


\begin{table*}[!htb]
\centering
\begin{tabular}{|c |c |c |c | c |  c| c| }
\hline
Model &   Slope &                Slope &                           Predicted& $c$   Estimate  & $c$  Estimate    \\  
    &   $99.73\%$ conf.    &  $95 \%$ conf.    &   Slope   &     $99.73\%$ conf.   &   $95\%$ conf.  \\ 
\hline \hline 
Lieb-Liniger &  $0.164^{+0.005}_{-0.005}$  & $0.164^{+0.003}_{-0.003}$ & $c/6 = 0.1666...$ &  $0.983^{+0.029}_{-0.030}$   & $0.983^{+0.019}_{-0.019}$  \\  \hline
Relativ. Boson &   $0.171^{+0.004}_{-0.004}$  & $0.1710^{+0.0022}_{-0.0022}$ & $c/6 = 0.1666...$ &  $1.026^{+0.021}_{-0.022}$   & $1.026^{+0.013}_{-0.013}$  \\ \hline
Quantum Ising &   $0.0826^{+0.0012}_{-0.0011}$  & $0.0826^{+0.0007}_{-0.0007}$ & $c/6 = 0.08333...$ &  $0.496^{+0.007}_{-0.007}$   & $0.496^{+0.004}_{-0.004}$  \\  
\hline
\end{tabular}
\caption{\footnotesize{ Summary of central charge estimates for the Lieb-Liniger, massless relativistic boson, and critical quantum Ising models obtained by scaling the entanglement entropy $S$ of a half-infinite line vs. $\log(\mu_2(D))$. }}
\label{tab:c_half_line_results}
\end{table*}

\begin{table*}[!htb]
\centering
\begin{tabular}{|c |c |c |c | c |  c| c| }
\hline
Model &  Slope &                Slope &                           Predicted& $c$   Estimate  & $c $ Estimate    \\  
    &    at $99.73\%$ conf.    &   $95 \%$ conf.    &   Slope   &    $99.73\%$ conf.   &  $95\%$ conf.  \\ 
\hline \hline 
Lieb-Liniger & $0.331^{+0.004}_{-0.004}$  & $0.3313^{+0.0026}_{-0.0027}$ & $c/3 = 0.333...$ &  $0.994^{+0.013}_{-0.013}$   & $0.994^{+0.008}_{-0.008}$  \\  \hline
Relativ. Boson &    $0.3365^{+0.0033}_{-0.0033}$  & $0.3365^{+0.0020}_{-0.0021}$ & $c/3 = 0.333...$ &  $1.010^{+0.010}_{-0.010}$   & $1.010^{+0.006}_{-0.006}$  \\ 
\hline
\end{tabular}
\caption{\footnotesize{ Summary of central charge estimates for the Lieb-Liniger and massless relativistic boson models obtained by scaling the entanglement entropy $S$ of an interval at scale $s=0.1$ vs. $\log(\mu_2(D))$.  The linearity of the fits is improved compared to those displayed in Table \ref{tab:c_half_line_results}}.    }
\label{tab:c_LL_results_interval}
\end{table*}

\begin{table*}[!htb]
\centering
\begin{tabular}{|c |c |c | c |  c | c |}
\hline
Operator   &  Optimal  &   $2\Delta_{\hatO}$ at $99.73\%$  & $2\Delta_{\hatO}$ at $95\%$ &  Exact result   \\
& scale  &   confidence & confidence   &  \\
\hline \hline 
$\hatpsi$ &   $0.66\mu_2$ & $0.1667^{+0.0005}_{-0.0005} $ &  $0.1665^{+0.0003}_{-0.0003}$ &  $0.1668575...$ \\ 
\hline
$ \frac{d}{ dx} \hatpsi $  &  $0. 86 \mu_2 $     & $2.165^{+0.006}_{-0.005}$     &$ 2.165^{+0.004}_{-0.003}$ &  $2.1668575...$        \\
\hline
$ \frac{d^2}{ dx^2} \hatpsi $  &   $1.49 \mu_2 $     & $4.167^{+0.010}_{-0.010}$     &$ 4.167^{+0.006}_{-0.007}$ &  $4.1668575...$        \\
\hline
$\hatpsidag \hatpsi$ & $0.965 \mu_2$ & $2.001^{+0.009}_{-0.008} $ & $2.001^{+0.005}_{-0.005} $ & 2 \\ 
\hline
$\hatcH$ & $1.58 \mu_2$ & $4.013^{+0.018}_{-0.019}$ &  $ 4.013^{+0.011}_{-0.013} $ & 4   \\ 
\hline
\end{tabular}
\caption{\footnotesize{ Summary of critical exponent estimates for the Lieb-Liniger model. } }
\label{tab:LL_results}
\end{table*}

\textbf{Field-field exponent ($\hatpsi-\hatpsidag$)}

The field-field exponent can be calculated using the  Bethe Ansatz to arbitrary precision.\cite{Korepin:1997kx} The general result reads:
\begin{align}
\label{eq:field-field}
\langle \hatpsi(x,t) \hatpsidag (0,0) \rangle \approx A | x + iv t |^{\frac{-1}{2 \calZ^2}} \ ,
\end{align}
where  $\calZ$ is given by
\begin{align}
Z(k) \equiv 2 \pi \rho(k)
\label{eq:bethe_Z}
\end{align}
evaluated at the Fermi-boundary of the quasi-momenta; $\rho$ is the density of quasi-momenta. For $g=1, v=1 \leftrightarrow g_{\mathrm{eff}} = 1.3478... $, the critical exponent is given by:
\begin{align}
\frac{1}{2 \calZ^2}  = 2 \Delta_{\hatpsi} = 0.1668575... \ .
\end{align}

We consider the correlator (\ref{eq:field-field}) at equal times, and restrict to $x>0$, so that the cMPS approximation is given by:
\begin{align}
\langle \hatpsi(x,0) \hatpsidag (0,0) \rangle \approx  ( l | (  1 \otimes \conjR )  e^{Tx}  ( R \otimes 1) |r)  \ .
\end{align}

The $U(1)$ symmetry of the exact  Lieb-Liniger ground state is broken by the cMPS approximation; the expectation value of the field,
\begin{align}
\langle \hatpsi \rangle \approx (l | R \otimes 1 | r)  \neq 0  \ ,
\end{align}
 scales to zero as $D$ is increased,  and the state approaches the true Lieb-Liniger vacuum, however convergence is very slow. 
In fact, the scaling of $\log( | (l | R \otimes 1 | r) |^2$ vs. $\log(\mu_2(D))$ yields a (sub-optimal) approximation for the critical exponent of $\hatpsi$, and corresponds to the dominant contribution to the scaling as $s\rightarrow \infty$. In Figure \ref{fig:LL_psi_disconn_correlator}  (Appendix \ref{app:LL_field_field}) one can see that, with the disconnected part included, power-law behaviour  is immediately evident for distances smaller then the correlation length, even for low bond dimension. This is not the case if the disconnected part is omitted.

\textbf{Descendants of $\hatpsi /  \hatpsidag$}

We examine the class of descendants of $\hatpsi$ at level $l$ obtained by taking  the $l$-th derivative of $\hatpsi$ (\ref{eq:psi_sq_cMPS}, \ref{eq:psi_cubed_cMPS}). While no exact Bethe Ansatz results are available for comparison, it follows from standard CFT arguments \cite{citeulike:1280772} that the exact exponent is simply $\Delta_{\frac{d^l}{ dx^l} \hatpsi} = \Delta_{\hatpsi} + l$, which is confirmed for first two levels to good accuracy (see Table \ref{tab:LL_results}).

\textbf{Density-density exponent ($\hatpsidag \hatpsi - \hatpsidag \hatpsi$)}

The Bethe Ansatz result  for the density-density correlator is:
\begin{align}
\label{eq:density-density}
& \langle \hatpsidag \hatpsi (x,t) \hatpsidag \hatpsi  (0,0) \rangle =   \langle \hatpsidag \hatpsi  \rangle^2 \\ \nonumber 
& +  \frac{A}{ (x + iv t )^{2}} +  \frac{A}{ (x - iv t )^{2}}  
+ A_3 \frac{ \cos (2 k_F x) }{| x + iv| ^{2 \calZ^2}} \ ,
\end{align}
where $A$ and $A_3$ are constants. Since $\calZ$ (\ref{eq:bethe_Z}) is bounded from below by $1$  \cite{Korepin:1997kx}, the first two terms dominate at large distances, so $\Delta_{\hatpsidag \hatpsi } = 1$. This is reproduced by our scaling calculations (see Table \ref{tab:LL_results}).

Unlike for the field-field correlator, here the disconnected part is non-zero in the exact ground state, so it needs to be subtracted out in the scaling calculation.

\textbf{ $\hatcH-\hatcH$ exponent}

The Hamiltonian density $\hat{\cH}$ is obtained from the time-time component of the energy-momentum tensor, which is a descendent of the unit operator.  For reasons equivalent to those given for the Hamiltonian density of the relativistic massless boson in the next section, $\delta_{\hatcH} = 2$, which our scaling calculation confirms (Table \ref{tab:LL_results}).

\subsection{Massless Relativistic Boson}
\label{subsection:KG}

Let us start from the massive relativistic boson (Klein-Gordon) Hamiltonian in (1+1) dimensions: 
\begin{align}
\label{eq:KG}
\hatH_{\mathrm{KG}} = \frac{1}{2}  \int_{-\infty}^{\infty} dx \left[ \hatpi^2 +  \left( \frac{d}{dx} \hatphi \right)^2 + m^2 \hatphi^2  \right]  \ .
\end{align}
For $m=0$ we obtain a conformally invariant theory with central charge $c=1$.  The field  operators $\hatphi$ and $\hatpi$ can be written in terms of the cMPS Fock space operators $\hatpsi$ and $\hatpsidag$ as:
\begin{align}
\hatphi = \frac{1}{\sqrt{2 \nu } } ( \hatpsi + \hatpsidag ) \ \ \ , \ \ \  \hatpi = -\frac{i}{2} \sqrt{2 \nu}  ( \hatpsi -  \hatpsidag ) \ ,
\end{align}
where an arbitrary scale $\nu$ is introduced. The Hamiltonian (\ref{eq:KG}) diverges in the cMPS setting and needs to be regularised. Surprisingly, one way to do this is by requiring the second derivative of $\hatpsi$ to be continuous. It is, however, difficult to impose such a constraint, and in any event this approach is too restrictive for our purposes since we actually want to work with operators that contain second order derivative terms. A better solution is to consider the counterterm:
\begin{align}
\label{eq:counterterm}
\frac{1}{\nu^2 } \left( \frac{d  \hatpi}{dx} \right)^2 \ ,
\end{align} 
which removes all divergences and serves as a momentum cutoff. The resulting Hamiltonian has the form:
\begin{align}
\label{eq:reg_KG}
\hatH = \int_{-\infty}^{\infty} dx \left[ \frac{d}{dx} \hatpsidag \frac{d}{dx} \hatpsi + v \hatpsidag \hatpsi + u ( \hatpsi \hatpsi + \hatpsidag\hatpsidag)  \right] 
\end{align}
with:
\begin{align}
v = \frac{m^2 + \nu^2}{2 }  \ \ \ , \ \ \  u = \frac{m^2 - \nu^2}{4  }  \ .
\label{eq:uu_parameters}
\end{align}


Results presented in this section are obtained using the values $u=-5$,  $v=10$.


Estimates for $\kappa$, together with the related estimates for $c$, are given in Table \ref{tab:kappa_results}. Estimates for the central charge obtained using the half-infinite line entropies, and from entropies of subsystems of length  $0.1 \mu_2(D)$, are summarised in Tables \ref{tab:c_half_line_results} and \ref{tab:c_LL_results_interval} respectively. The comments made in the context of the Lieb-Liniger model regarding the accuracy of the value for $\kappa$ as predicted by (\ref{eq:kappa_def}), and the scaling of the transfer matrix eigenvalues, apply here as well.


Critical exponent estimates are listed in Tables \ref{tab:uu_results} and  \ref{tab:vertex_results}; the latter lists estimates for vertex operator $: \exp(i \beta \hatphi): $ exponents, for a range of values for the free parameter $\beta$.

\begin{table*}[!htb]
\centering

\begin{tabular}{|c |c |c | c |  c | c |}
\hline
Operator   & Optimal  &   $2\Delta_{\hatO}$ at $99.73\%$  & $2\Delta_{\hatO}$ at $95\%$ &  Exact result   \\
 & scale  &   confidence & confidence   &  \\
\hline \hline 
$\partial_z \hatphi$ &   $ 0.25 \mu_2$ & $2.00013^{+0.00028}_{-0.00027} $ & $2.00013^{+0.00017}_{-0.00016} $ & 2 \\ 
\hline
$ \partial_z^2 \hatphi $     &  $1.63  \mu_2 $     & $3.992^{+0.008}_{-0.009}$     &$ 3.992^{+0.006}_{-0.006} $ &  $4$       \\
\hline
$ \partial_z^3 \hatphi $  &  $3.96 \mu_2 $     & $6.007^{+0.005}_{-0.006}$     &$ 6.001^{+0.003}_{-0.004} $ &  $6$       \\
\hline
$\hatcH$ &  $0.78 \mu_2$ & $3.97^{+0.06}_{-0.07}$ &  $ 3.97^{+0.03}_{-0.05}$ & 4   \\ 
\hline
\end{tabular}
\caption{\footnotesize{Summary of critical exponent estimates for the massless relativistic boson.}}
\label{tab:uu_results}
\end{table*}

\textbf{ $\partial_z \hatphi$ exponent}

$\partial_z \hatphi$ is a $(2,0)$ primary field, so  $\Delta = 1$. Our scaling calculation reproduces this to remarkable accuracy (see Table \ref{tab:uu_results}).  The relevant expression in terms of cMPS creation and annihilation operators is obtained as follows. Performing a Wick rotation back to Minkowski space,  we have $\partial_z = \frac{1}{2} ( \partial_x - \partial_t )$, so:
 \begin{align}
 \partial_z \hatphi = & \frac{1}{2} \left( \frac{d}{dx} \hatphi - \hatpi \right) \\ \nonumber   = &  \frac{1}{2}  \left[ \frac{1}{\sqrt{2 \nu} } \frac{d}{dx}( \hatpsi + \hatpsidag) +  \frac{i \sqrt{2 \nu}}{2} ( \hatpsi - \hatpsidag) \right] \ .
 \end{align}

\textbf{Descendants of $\partial_z \hatphi$ } 

In order to obtain the expression without time derivatives, which is necessary in order to write down the correlator in terms of cMPS data,  we first start by expanding, 
\begin{align}
\label{eq:dtdtphi}
\partial_z \partial_z \hatphi = \frac{1}{4} \left( \frac{d^2}{dx^2} \hatphi - 2 \frac{d}{dx} \hatpi  + \frac{d}{dt}\hatpi \right) \ ,
\end{align}
and next use:
\begin{align}
\frac{d}{dt} \hatpi = \frac{\delta \hatH}{\delta \hatphi} = -  \frac{d^2}{dx^2} \ \hatphi \ .
\end{align}
 The final result is simply:
\begin{align}
\partial_z \partial_z \hatphi = - \frac{1}{2} \frac{d}{dx} \hatpi  \ .
\end{align}
The time derivative of the canonical momentum in (\ref{eq:dtdtphi}) precisely cancels the double spatial derivative of $\hatphi$. It should be noted that a $\delta$-function divergence occurs in cMPS expectation values when two operators containing second and higher order spatial derivatives coincide. This is not a problem in the present context since we are not interested in taking the limit in which two operators are at exactly the same position. Second order (and higher) spatial derivatives of  $\hatpsi / \hatpsidag$  (\ref{eq:psi_sq_cMPS}, \ref{eq:psi_cubed_cMPS}) are present in cMPS expressions when evaluating $(\partial_z)^n \hatphi$ for $n>2$.

The above approach for eliminating time derivatives can be applied straightforwardly for an arbitrary number of $\partial_z$ derivatives. Each application of $\partial_z$ increases the value of the critical exponent by one. The numerical results for descendants up to the third level are displayed in Table \ref{tab:uu_results}.

\textbf{Energy-momentum tensor and Hamiltonian density exponent}

The operator product expansion for the energy-momentum tensor $\mathbf{\hatT}$ \cite{citeulike:1280772},
\begin{align}
\mathbf{\hatT}_{z z} = : \partial_z \hatphi \partial_z \hatphi : \ ,
\end{align}
 with itself is given by:
\begin{align}
\label{eq:TT_ope}
\bold{\hatT}_{z z}  (z) \bold{\hatT}_{z z}  (0) = &   \frac{c (\alpha ')^2}{2 z^4}  - \frac{2 \alpha'}{z^2} \bold{\hatT}_{z z} (0)  \\ \nonumber & - \frac{2 \alpha'}{z} : \partial_z^2 \hatphi \partial_z \hatphi (0) : \ .
\end{align}
In our conventions $\alpha' = \frac{1}{2 \pi}$.  

The Hamiltonian density is simply the combination:
\begin{align}
\label{eq:H_in_terms_of_T}
\bold{\hatT}_{z z} + \bold{\hatT}_{\overline{z} \overline{z}}  = \hatcH \ ,
\end{align}
The appropriate OPE follows straightforwardly from (\ref{eq:TT_ope}), since the OPE of mixed $zz$ and $\overline{z} \overline{z}$ terms vanishes. Furthermore,  the second and third terms on the RHS in the OPE (\ref{eq:TT_ope}) drop out in the vacuum expectation value when considering only the connected component of the $\hatcH$ - $\hatcH$ correlator.

In conclusion, only the first term in (\ref{eq:TT_ope}) survives in the vacuum expectation value,  so $\Delta_{\hatcH} = 2$, which is reproduced by our numerics (see Table \ref{tab:uu_results}).

\textbf{Vertex Operators}

The free relativistic massless boson CFT has an infinite number of primary operators of the form $: \exp(i \beta \hatphi): $ (where  $ : :$ denotes normal ordering), parameterised by a  real coefficient $\beta$. The scaling exponent for each such operator is:
\begin{align}
2 \Delta = \frac{\alpha ' \beta^2}{2} = \frac{\beta^2}{2 \pi} \ ,
\end{align}
where the last equality assumes our conventions.

The cMPS approximation is given by:
\begin{align}
 \langle 0 | : \exp(i \beta \hatphi):  & \cdots  | \rangle  \approx \\ \nonumber 
& (l | \exp \left( \frac{i \beta}{\sqrt{\nu}} (R \otimes 1 + 1 \otimes \overline{R}) \right) \cdots | r)  \ ,
\end{align}
where $\cdots$ denotes additional insertions. Critical exponent estimates for a range of values for $\beta$ are displayed in Table \ref{tab:vertex_results}.

\begin{table*}[!htb]
\centering
\begin{tabular}{|c |c |c | c |  c|  }
\hline
$\beta$   & Optimal & $2\Delta_{\hatO}$ at $99.73\%$ & $2\Delta_{\hatO}$ at $95 \%$  & Exact result  \\
		& scale    & confidence                                  &      confidence                          &                   \\  
\hline \hline 
$0.1$ &  $0.81 \mu_2$ & $(1.589^{+0.012}_{-0.012}) \e{-3}$ & $(1.589^{+0.008}_{-0.008}  )\e{-3}$ &  $1.592... \e{-3}$ \\ 
\hline
$0.2$ &  $0.83 \mu_2$ & $(6.36^{+0.05}_{-0.05}) \e{-3}$ & $(6.36^{+0.03}_{-0.03}) \e{-3}$ &  $6.366... \e{-3}$  \\ 
\hline
$0.4$ &  $0.90 \mu_2$ & $(2.547^{+0.023}_{-0.023} ) \e{-2}$ & $(2.547^{+0.014}_{-0.015} ) \e{-2}$ &  $2.546... \e{-2}$  \\ 
\hline
$0.6$ &  $0.98 \mu_2$ & $(5.74^{+0.06}_{-0.06} ) \e{-2}$ & $(5.74^{+0.04}_{-0.04}) \e{-2}$ &  $5.792... \e{-2}$  \\ 
\hline
$1$ &  $1\mu_2$ & $0.1595^{+0.0016}_{-0.0017} $ & $0.1595^{+0.0010}_{-0.0011} $ &  $0.1591... $  \\ 
\hline
$2$ &  $1 \mu_2$ & $0.637^{+0.007}_{-0.007} $ & $0.637^{+0.005}_{-0.004}$ &  $0.6366...$  \\ 
\hline
$3$ &  $1\mu_2$ & $1.433^{+0.022}_{-0.022}$ & $1.433^{+0.014}_{-0.014}$ &  $1.432...$  \\ 
\hline
\end{tabular}
\caption{\footnotesize{ Summary of critical exponent estimates for the vertex operator $:\exp (i \beta \hatphi):$ for a range of values for the parameter $\beta$. The accuracy starts to degenerate abruptly beyond $\beta \approx 3$ due to ultra-violet cutoff effects.}}
\label{tab:vertex_results}
\end{table*}
Due to  the presence of a finite cutoff $\nu$, the FES scaling algorithm eventually fails  to reproduce the exponents as the value of $\beta$ is increased; indeed beyond  $\beta \approx 3$ the estimates degenerate quickly.

\subsection{Quantum Ising Model}

The Hamiltonian of the quantum Ising model in a transverse magnetic field on an infinite $1d$ chain is given by:
\begin{align}
\hatH = \sum_{i \in \mathbb{Z}} -J \hatsigma^x_i \hatsigma^x_{i+1} + h \hatsigma^z_i \ ,
\label{eq:ising_hamiltonian}
\end{align}
where $\{ \hatsigma^x, \hatsigma^y, \hatsigma^z \}$ are the Pauli matrices, $J$ determines the coupling strength between nearest neighbour spins, and $h$ determines the strength of the magnetic field.  The model is critical for $h/J = \pm 1$. The numerics in this section are performed using $J=-1$ and $h=1$, and a spline interpolation is used in order to obtain  values for two-point correlation functions at arbitrary distances (see footnote on page \pageref{ftn:interpolation}).

The quantum Ising model can be mapped to a free fermion model and solved exactly; the CFT describing the theory at the critical points  $h/J = \pm 1$ has central charge $c=1/2$. 

The low lying eigenvalues of the transfer matrix can be seen to all scale in the same way, their ratios converging to definite values as the bond dimension is increased. This is depicted in the plots in Figure \ref{fig:Ising_combined}. For a theoretical interpretation of this convergence see the discussion at the beginning of Section \ref{sec:D_scaling}, 

Estimates for the central charge are presented in Tables  \ref{tab:c_half_line_results} and \ref{tab:c_LL_results_interval}.  The estimate for $\kappa$ is given in Table \ref{tab:kappa_results} in Appendix \ref{app:kappa}, and the relevant comments made in the context of the Lieb-Liniger model apply here as well.

Since the underlying CFT describing the critical quantum Ising model is minimal, it has a finite number of primary fields. \cite{citeulike:1280772} There are five in total - two correspond to local and three to non-local operators.   The two local primaries are traditionally denoted as $\hatsigma$ and $\hatepsilon$, and using our conventions  (\ref{eq:ising_hamiltonian}) they are given by:
\begin{align}
\hatsigma(i) = \hatsigma^x_i \ \ \ , \ \ \  \hatepsilon(i) = \hatsigma^x_i \hatsigma^x_{i+1} - \hatsigma^z_i \ .
\end{align}
The three non-local primates are denoted as $\hatmu$, $\hatpsi$, and $\hatpsibar$. $\hatmu$ is given by a half-infinite string consisting of $\hatsigma^z$-s up to (and including) position $i$, while  $\hatpsi$ and $\hatpsibar$ have instead $\hatsigma^+ := \frac{1}{2} (\hatsigma^x + i \hatsigma^y)$ and $\hatsigma^- := \frac{1}{2}  (\hatsigma^x - i \hatsigma^y)$ at position $i$. These strings modify the MPS transfer matrix but otherwise do not change our method for extracting the corresponding critical exponents.

We also consider a class of descendant fields obtained by taking discrete derivatives of the local primaries; for example the first level descendant of $\hatsigma$ is $d \hatsigma(i)  := \hatsigma(i+1) - \hatsigma(i)$.

The estimates for the critical exponents are displayed in Table \ref{tab:ising_results}, which also contains the exact values. We note that for many operators there is no clear intersection between the Direct and FES Approaches (see Section \ref{sec:D_scaling}), so when this is the case we need to work with $s_0 =1$ in our algorithm, i.e. we perform the scan over scale from $s=1$ to $s=\infty$. 

\begin{table*}[!htb]
\centering

\begin{tabular}{|c |c |c | c |  c | c |}
\hline
Operator   & Optimal  &   $2\Delta_{\hatO}$ at $99.73\%$  & $2\Delta_{\hatO}$ at $95\%$ &  Exact result   \\
& scale  &   confidence & confidence   &  \\
\hline \hline 
$\hatsigma$ &  $1 \mu_2$ & $0.2492^{+0.0008}_{-0.0010}$ &  $0.2492^{+0.0005}_{-0.0006} $ &  0.25  \\ 
\hline
$d \hatsigma$ &  $ 1.25 \mu_2$ & $2.250^{+0.003}_{-0.004} $ & $2.2497^{+0.0021}_{-0.0020} $ & 2.25 \\ 
\hline
$ d^2 \hatsigma $  &  $2.15  \mu_2 $     & $4.248^{+0.006}_{-0.006}$     &$ 4.248^{+0.004}_{-0.004} $ &  $4.25$       \\
\hline
$ d^3 \hatsigma $  &  $3.2  \mu_2 $     & $6.249^{+0.008}_{-0.008}$     &$ 6.249^{+0.005}_{-0.005} $ &  $6.25$       \\
\hline
$\hatepsilon$ & $4 \mu_2$ & $1.996^{+0.005}_{-0.005}$ &  $ 1.996^{+0.003}_{-0.003}$ & 2   \\ 
\hline
$d \hatepsilon$ &  $1.85 \mu_2$ & $3.997^{+0.010}_{-0.010}$ &  $ 3.997^{+0.007}_{-0.007}$ & 4   \\ 
\hline
$ \hatmu $ & $\infty \mu_2$ & $0.2508^{+0.0018}_{-0.0017}$ &  $ 0.2508^{+0.0011}_{-0.0010}$ & 0.25   \\ 
\hline
$\hatpsi / \hatpsibar$ &  $1.95 \mu_2$ & $0.9991^{+0.0013}_{-0.0013}$ &  $ 0.9991^{+0.0008}_{-0.0008}$ & 1   \\ 
\hline
\end{tabular}
\caption{\footnotesize{Summary of estimates for the critical exponents of the critical quantum Ising model.}}
\label{tab:ising_results}
\end{table*}

\section{Conclusions}
\label{sec:conclusions}

In this paper we have developed finite entanglement scaling (FES) methods, based on translation invariant (continuous) matrix product states in the thermodynamic limit, for calculating conformal field theory (CFT) data for critical theories, namely the central charge and critical exponents of both local and nonlocal scaling operators. The fact that for the three exemplary models our algorithm is capable of reproducing the exact results to excellent accuracy using only a modest range of bond dimensions provides strong support for the validity of the FES hypothesis  (\ref{eq:entropy_scaling_general}, \ref{eq:2pcorr_scaling_general}) presented in Section \ref{sec:scalinghypothesis}. One of the new ingredients in our approach is to directly use the (c)MPS induced correlation length as scaling parameter, rather than the bond dimension or any function thereof. This is essential to obtain the accuracy on the data reported in this paper. The calculation of operator product coefficients between primary fields, has not been addressed in this paper. This involves a three-point function scaling calculation and will be addressed in a future publication. Together with the central charge and critical exponents of the primaries, the operator product coefficients constitute the data necessary to fully specify a general (i.e. non-minimal) CFT.\cite{citeulike:1280772}

Crucial to the precision is the ability to optimise over the scale parameter $s$ at which critical exponents are calculated. This optimisation hinges on the fact that it is not only the first eigenvalue of the transfer matrix that scales with $D$ as $D^{\kappa}$, but all the other low-lying eigenvalues also follow the same scaling. This results was not presented before and provides a further hint that there should exist a CFT interpretation for finite entanglement scaling, that once fully understood would provide access to the sub-leading corrections and possibly to a geometric interpretation of FES.

The FES calculations have been performed for three exemplary models: two field theories, the (non-relativistic)  Lieb-Liniger model and the massless relativistic boson, and to the critical quantum Ising model in the lattice setting. The numerical accuracy of the results is comparable to those of MERA calculations.\cite{2008PhRvL.101k0501V,2010PhRvB..82m2411E,2011arXiv1109.5334E} The central advantage over MERA is the computational cost, which is much lower for comparable accuracy. In addition, the continuous version of MPS can equally be applied to free and interacting field theories, while there is no interacting version of the continuous version of the MERA as of yet.\cite{2011arXiv1102.5524H}
The central disadvantages include the fact that at present a geometric or a renormalisation group interpretation of the CFT perturbation caused by the finite bond dimension is lacking, and the related problem that we do not understand how the structure of the CFT is encoded in the (c)MPS data. What we mean by the latter is some mapping between the primary and descendant structure of the CFT and the eigen-decomposition of the (c)MPS transfer matrix -  a practical benefit of such a mapping would be that we could simply work at the level of (c)MPS, without needing any additional information about the primary/descendent structure in terms of operators acting at the physical level. There has been some progress in the MPS context along these lines for the entanglement spectrum\cite{2013arXiv1303.0741L}, albeit not in the thermodynamic limit. We are hoping to report on some new findings in this direction soon. In addition, it would also be interesting to check wether the finite entanglement scaling framework can be used for determining critical exponents of boundary CFTs corresponding to edges in the system, analogous to the MERA results presented in Ref.~\onlinecite{2010PhRvB..82p1107E}.

Finally, let us turn to the issue of determining the critical point. The models studied in this paper either have an extended critical region, or have a critical point whose location is known exactly\footnote{The $\phi^4$ model with an imaginary mass parameter is an interesting theory for which this is not the case (see \cite{2013PhRvD..88h5030M} for a MPS based study of critical regions in this model).}: the Lieb-Liniger model is critical for all choices of parameters in (\ref{eq:LL_hamiltonian}), the relativistic boson model (\ref{eq:reg_KG}, \ref{eq:uu_parameters}) is critical for $m =0$, while the transverse quantum Ising model (\ref{eq:ising_hamiltonian}) is critical for  $h/J = \pm 1$. When the values for the parameters at criticality are not known, one can try to obtain them from the (c)MPS simulation. Let us illustrate this in the context of the quantum Ising model, by imagining that, having chosen e.g. $J=1$, we do not know that the critical point is at $h=1$. In order to obtain an estimate it is necessary to first scan over $h$ for a range of bond dimensions and search for the point at which the order parameter $\langle \hatsigma^x \rangle$ transitions from a finite value to zero. For finite bond dimension this happens at some point $h(D)>1$ and the exact critical point $h=1$ can be obtained by scaling to $D \rightarrow \infty$ \cite{2008PhRvB..78b4410T} (see the plot in Figure \ref{fig:Ising_phase_transition}). This raises two questions. Firstly, one can wonder how sensitive the results are to the accuracy with which the exact critical point $h(D\to\infty)$ is obtained. Secondly, one can question whether it may be more natural to perform the scaling calculations using (c)MPS solutions obtained at the transition point $h(D)$ at each bond dimension $D$, rather than using the exact point $h(D\to\infty)$. We can answer the second question negatively. Both for the quantum Ising model and in a preliminary cMPS analysis of the $\phi^4$ model \cite{phi4_in_preparation}, we have established that the FES scaling approach does not work ---or needs to be altered--- when using the (c)MPS transition points. To directly extract the scaling exponents of the primary operators, the Hamiltonian parameters have to be kept fixed. Regarding the first question, we anticipate that by using the (c)MPS induced correlation length, the scaling hypothesis of Eq.~\eqref{eq:entropy_scaling_general} and \eqref{eq:2pcorr_scaling_general} continues to hold as long as the parameters of the Hamiltonian are sufficiently close to the critical point so that we are in the scaling regime. Even when the bond dimension grows sufficiently large so as to accurately reproduce the slightly off-critical ground state, this will only cause a saturation in $\mu_2(D)$ so that no new data points are obtained by further increasing $D$. At this point, the scaling relation $\mu_2(D)\sim D^{\kappa}$ will break down, which is why the use of $\mu_2(D)$ as scaling parameter is to be preferred.

\begin{figure}[!htb]
\centering
\includegraphics[width=8cm]{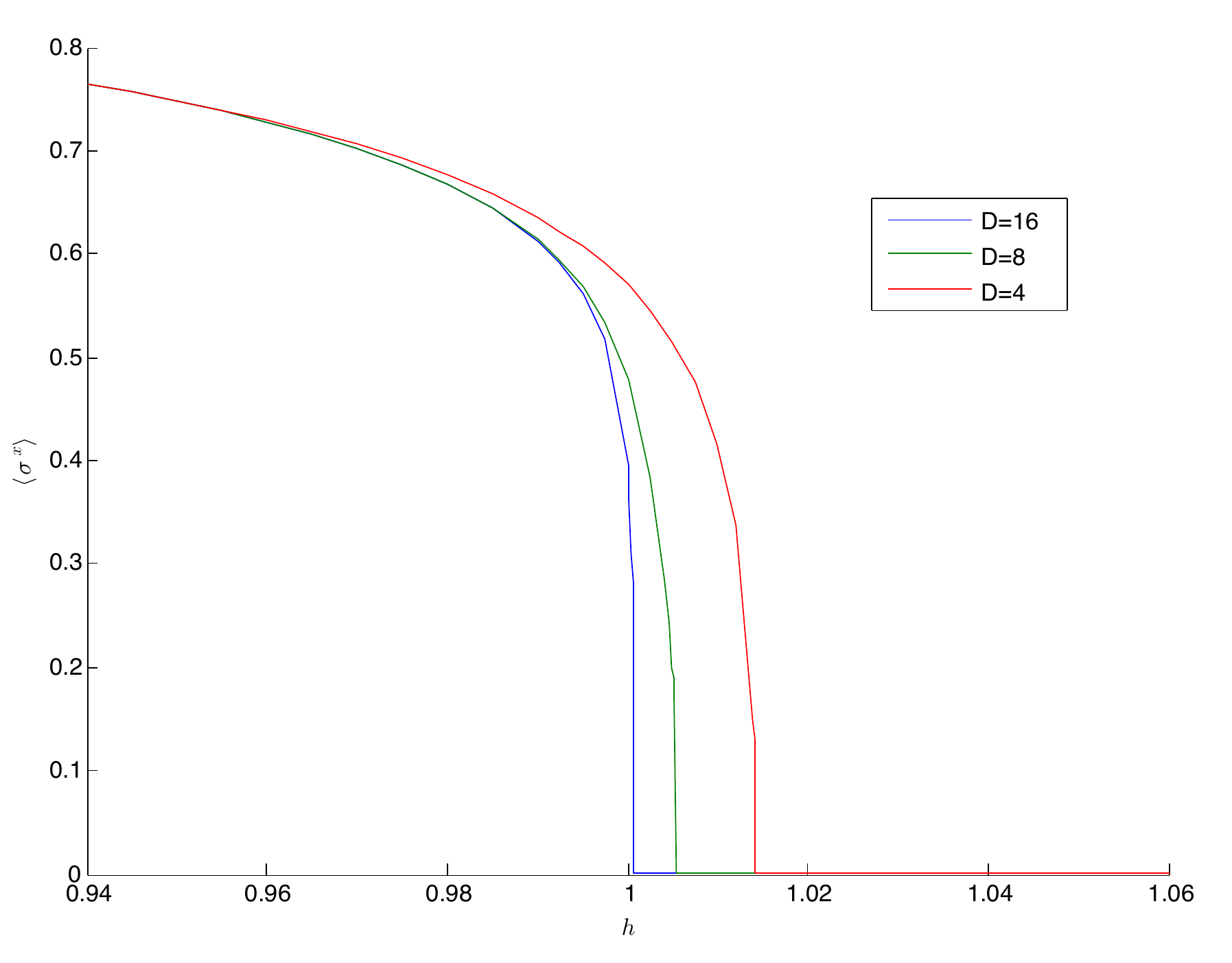}
\caption{\footnotesize{ (Colour online). The order parameter of the Ising model $\langle \hatsigma^x \rangle$ as a function of magnetic field strength parameter $h$ at bond dimensions $D=4,8,16$. The order parameter transitions abruptly from a positive value to zero at a point that is slightly larger than the exact value $h=1$. The transition point approaches the exact value with increasing $D$. } }\label{fig:Ising_phase_transition}
\end{figure}


\begin{acknowledgements}
We would like to thank  Marek Rams, Volkher Scholz, Henri Verschelde, and Valentin Zauner for helpful discussions.
\end{acknowledgements}

\appendix

\section{Review of Continuous Matrix Product States}
\label{app:cMPS_review}

The variational set of matrix product states (MPS) is given by:
\begin{align}
& \ket{ \Psi [ A ] }= \\ \nonumber & \sum_{i_1=1}^d \sum_{i_2=1}^d \cdots \sum_{i_N=1}^d v_L^\dagger A^{i_1}_1 A^{i_2}_2 \cdots A^{i_N}_N  v_R \ket{ i_1, i_2, \cdots, i_N }  \ ,
\end{align}
where $d$  is the number of physical (spin) degrees of freedom, and for every value of the index $i_a$, $A^{i_a}$ is a $D\times D$ matrix. In order to take the continuum limit we first promote the finite-dimensional Hilbert space at each lattice site to a full Fock space:
\begin{align}
& \hata_i \ket{\Omega} = 0  \ \ \  , \ \ \   [ \hata_i, \hata_j ] = 0 = [ \hata^\dagger_i, \hata^\dagger_j  ]  \\ \nonumber 
&   [ \hata_i, \hata_j^\dagger ] = \delta_{ij}  \ .
\end{align}

The continuum limit $\epsilon \rightarrow 0$ is taken as \cite{2012arXiv1211.3935H}:
\begin{align}
& \ket{ \Psi [ A ] }_\epsilon= \\ \nonumber  & \sum_{i_1 \cdots i_N}^d \left(  v_L^\dagger A^{i_1}_{\left(-\frac{N}{2} \right)}  \cdots A^{i_N}_{\left( \frac{N}{2} \right)}  v_R  \right) (\psihdag_1)^{i_1} \cdots (\psihdag_N)^{i_N} \ket{ \Omega}   \ ,
\end{align}
where
\begin{align}
\psihdag_i = \frac{\hata_i^\dagger}{\sqrt{\epsilon}} \ \ \  \psih_i = \frac{\hata_i}{\sqrt{\epsilon}}   \ \ \ , \ \ \   N = \frac{L}{\epsilon} \ .
\end{align}
This limit can be taken consistently only if the infinite set of matrices $A^i$ depends on two matrices $R$ and $Q$ as:
\begin{align}
A^0 = \eye + Q   \  \ \  , \ \ \  A^1 = \epsilon R \ \ \ ,  \ \ \  A^n = \epsilon^n \frac{R^n}{n!}  \ .
\end{align}
Promoting the above analysis to multiple particle species, the continuous matrix product variational set of states (cMPS) on a finite interval $[ -L/2, L/2 ]$, can be written as:
\begin{align}
\label{eq:cMPS}
& \ket{ \Psi [ Q(x), R_{\alpha}(x) ] } = \\ \nonumber &  v_L^\dagger \mathcal{P} \mathrm{exp} \left[  \int_{- \frac{L}{2}}^{\frac{L}{2}} dx   \ \left(Q(x) \otimes \eye + \sum_\alpha R_{\alpha} \otimes \hatpsidag_\alpha (x)  \right)  \right]  v_R  \ket{ \Omega}   \ .
\end{align}
The $\alpha$ index runs over  particle species, $\mathcal{P} \mathrm{exp}$ denotes the path ordered exponential,  and $v_L$, $v_R$, determine the boundary conditions. If the particles are bosons $[\psi_\alpha(x) , \psidag_\beta (y) ]  = \delta_{\alpha \beta} (x -y) $, while for fermions $\{ \psi_\alpha(x) , \psidag_\beta (y) \}  = \delta_{\alpha \beta} (x -y) $.

In this paper we are interested in translation invariant cMPS describing  a single bosonic  particle species in the thermodynamic limit, that is, the variational set:
\begin{align}
\label{eq:ucMPS}
& \ket{ \Psi [ Q, R ] } \\ \nonumber 
& = v_L^\dagger \mathcal{P} \mathrm{exp}  \left[ \int_{- \infty}^{\infty} dx   \ \left( Q \otimes \eye +  R \otimes \hatpsidag  \right)  \right]  v_R \ket{ \Omega}   \ ,
\end{align}
with the matrices $R$ and $Q$ position independent. The transfer matrix is given by
\begin{align}
\label{eq:transfer_matrix}
T = Q \otimes \eye  + \eye \otimes \overline{Q} + R \otimes \overline{R} \ .
\end{align}
Finite normalisation requires the largest eigenvalue of the transfer matrix to be zero, which can always be achieved by transforming $Q \rightarrow Q - (\lambda / 2) \eye$, where $\lambda$ is the initial largest non-zero eigenvalue of $T$. 

In this paper we  find it convenient to define the transfer matrix for MPS in the thermodynamic to be:
\begin{align}
T_{\mathrm{MPS}} = \log(E) \ ,
\label{eq:MPS_transfer_matrix}
\end{align}
 where
\begin{align}
E = \sum_i A^i \otimes \overline{A}^i   \ .
\end{align}
Usually $E$ itself is referred to as the transfer matrix in MPS literature, but  as this is inconsistent with the cMPS conventions, we chose to define $T$ as in (\ref{eq:MPS_transfer_matrix})  instead.



Expectation values involving an insertion of a single operator involve only the left and right zero-eigenvalue eigenvectors $(l |$, and $|r)$.  We normalise these so that the state has norm one:
\begin{align}
\braket{\Psi}{ \Psi}  = (l | r)  = 1 \ .
\end{align}
Expectation values of insertions of $\hatpsi$, $\hatpsidag$ have straightforward cMPS expressions, for example:
\begin{align}
& \bra{\Psi} \hatpsi \ket{\Psi} = ( l | R \otimes \eye |r ) = \mathrm{tr} ( l^{T} R  r)  \\ \nonumber
&  \bra{\Psi} \hatpsidag \ket{\Psi} = ( l | \eye \otimes \Rbar |r ) = \mathrm{tr} ( l^{T} r \Rdag )  \\ \nonumber
&  \bra{\Psi} \hatpsidag \hatpsi \ket{\Psi} = ( l | R \otimes \Rbar |r ) = \mathrm{tr} ( l^{T} R  r \Rdag )  \\ \nonumber
&  \bra{\Psi} \frac{d \hatpsi}{d x}  \ket{\Psi} =  ( l | [Q, R ] \otimes \eye |r ) = \mathrm{tr} ( l^{T} [Q, R]  r)  \ .
\end{align}
$l$ and $r$ in the rightmost expressions denote $D \times D$ matrices corresponding to the  $D^2$ component co-vector $(l|$ and vector $|r)$. Working with the trace expressions rather than the tenors product ones is clearly computationally more efficient, as  it involves manipulating $D \times D$ rather than $D^2 \times D^2$ matrices (computational cost $\cO (D^3)$ vs. $\cO (D^6)$ ). 

It is straightforward but tedious to calculate expressions involving higher derivatives of $\hatpsi$, \cite{2012arXiv1211.3935H} which we frequently require in this paper. In particular, the cMPS expression are more complicated than the expression for  $\frac{d \hatpsi}{d x} $  above suggests, and do not consist simply of a Kroenecker product of a nested commutator with the identity operator in $D$ dimensions. For example:
\begin{align}
\label{eq:psi_sq_cMPS}
& \bra{\Psi} \frac{d^2 \hatpsi}{d x^2}  \ket{\Psi} =  \\ \nonumber 
&   \ \ \ \ \ ( l | \left( [Q, [Q, R ] ] \otimes \eye  +  [R, [Q, R] ] \otimes \overline {R}\right) |r )   \ ,
\end{align}
and
\begin{align}
\label{eq:psi_cubed_cMPS}
 \bra{\Psi} \frac{d^3 \hatpsi}{d x^3}  \ket{\Psi} =   &
( l |   \left( \vphan [Q, [Q, [Q, R ] ] ] \otimes \eye  \right. \\ \nonumber 
& \ \ \ \ \  +  2 [R, [Q, [Q, R] ]  ] \otimes \overline {R}  \\ \nonumber
& \ \ \ \ \  \left. +    [R, [Q, R] ]   \otimes [ \overline{Q}, \overline{R} ]   \vphan   \right) |r )   \ .
\end{align}

Expectation values of operators at different spatial points separated by some finite distance $(x-y)$ involve the full transfer matrix. For example:
\begin{align}
& \bra{\Psi} \hatpsidag(x)  \hatpsi (y) \ket{\Psi} =  \\ \nonumber  &  \ \ \ \ \  ( l  | ( \eye \otimes \Rbar )     \exp \left[ T (y-x) \right] (R \otimes \eye ) |r ) \ \ \   \ \ \ y > x \\ \nonumber
&  \ \ \ \ \  ( l  | ( R \otimes \eye )     \exp \left[ T (x-y) \right] (\eye \otimes \Rbar ) |r ) \ \ \  \ \ \ x > y  \ ,
\end{align}
so unless $(x-y)$ is much larger than the correlation length, all the eigenvalues of the transfer matrix contribute. The above expressions can still be computed in $\cO (D^3)$, by exploiting the tensor product structure of the expressions to calculate the initial density matrix, (e.g. $( l  | ( \eye \otimes \Rbar ) $ for $x>y$ in the above example, which can be obtained at cost $\cO (D^3)$ ), and then using a partial differential equation solver to calculate the action of $\exp(T (y-x ) )$ on this co-vector.

\section{Details of the Lieb-Liniger Field-Field Exponent Calculation}
\label{app:LL_field_field}

In this section we describe the details of  the finite entanglement scaling (FES) approach for calculating critical exponents, using the example of the field-field ($\hatpsi-\hatpsidag$) correlator in the Lieb-Liniger model.  The algorithm, as described in  Section \ref{sec:D_scaling}, is to apply the FES Approach, aided by the Direct Approach; the role of the latter is simply to provide an estimate for the lower bound when scanning over scales in the FES Approach. 
\begin{figure}[!htb]
\centering
\includegraphics[width=8cm]{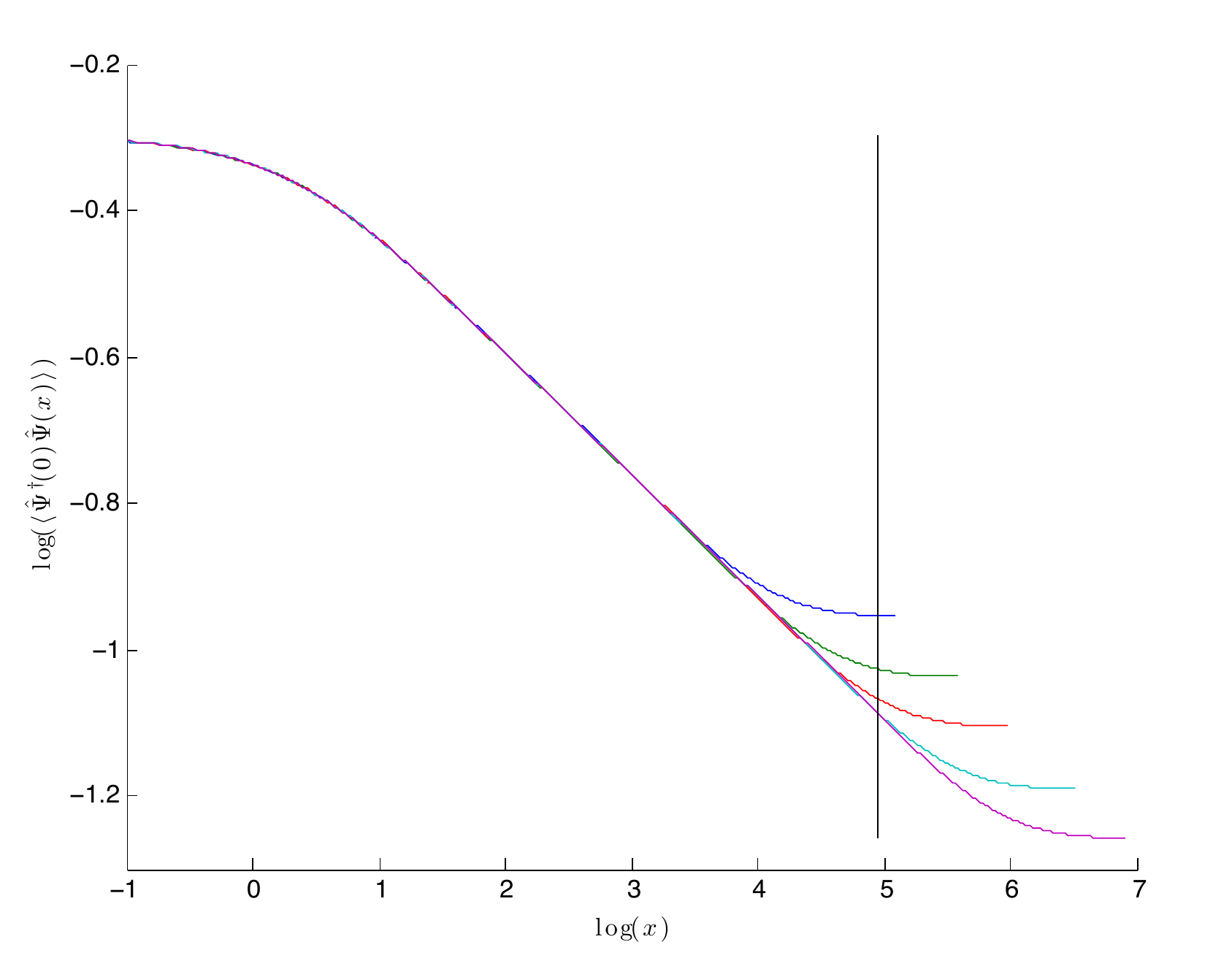}
\caption{\footnotesize{(Colour online). cMPS approximations for  $\log(\langle \hatpsi(x,0) \hatpsidag (0,0) \rangle )$ as a function of  $\log(x)$, with the connected  part $| ( l | (  1 \otimes \conjR ) | r ) |^2$ included.  Power law decay is approximated well up to the correlation length, beyond which exponential decay takes over. Results for bond dimensions $D=16,24,32,48,64$ are displayed (for smaller $D$ deviation from linear behaviour occurs at a shorter distance).  The vertical line denotes the correlation length at $D=64$. After rescaling the units of distance, for each of the above values of $D$, the  above curves correspond to slices through the surface displayed in Figure \ref{fig:LL_field_plot3d} at $\log(\mu_2) = \log(\mu_2(D))$.}}
  \label{fig:LL_psi_disconn_correlator} 
  \end{figure} 
By comparing with the known exact value for the exponent, the left plot in Figure \ref{fig:LL_field_scaling_direct_combined} demonstrates that FES gives good estimates for scales from infinity down to around $s=0.5$. Below this the linearity of the interpolation improves further, but the estimates are off due to short distance effects. The best estimate is roughly around $s=0.6$.  

 \begin{figure}[!htb]
\centering
\includegraphics[width=8cm]{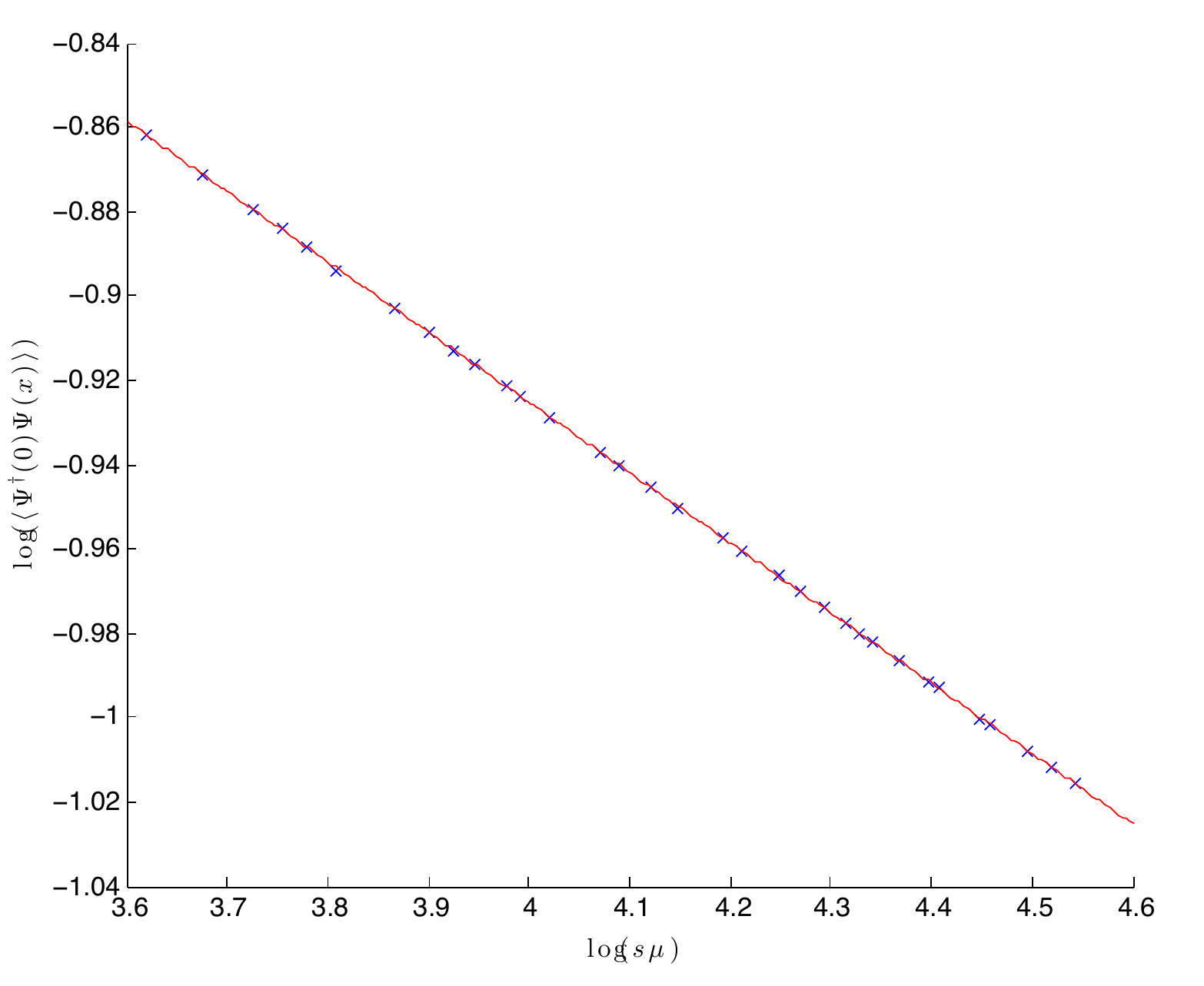}
\caption{\footnotesize{(Colour online).  The scaling of $\log(G(s \mu_2(D))$ vs. $\log( \mu_2(D))$ for $s=0.66$. The red line denotes the result of the interpolation, and corresponds to a slice of the two-dimensional surface in Figure \ref{fig:LL_field_plot3d}} through $\log(s) = \log(0.66)$.}
\label{fig:LL_field_logcorr_vs_logpos_with_fit}
\end{figure}

 The problem is that we do not a priori know the value for $s$ below which short distance effects destroy the precision of the FES scaling. The simplest solution is to simply pick the safe value $s=1$, which in itself is not a bad option as it significantly improves the accuracy over that  obtained at $s = \infty$.  In order to do better than this, we combine the FES and Direct Approach. 
 \begin{figure*}[!htb]
\centering
\includegraphics[width=16cm]{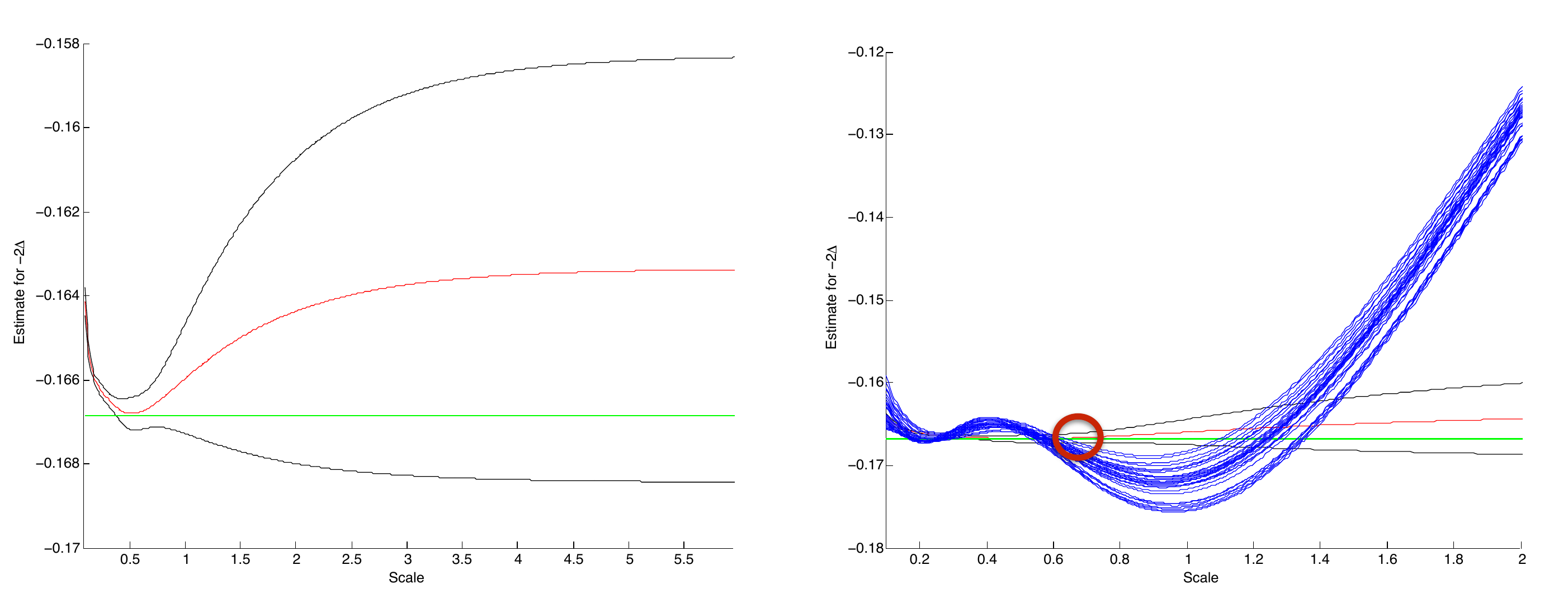}
\caption{\footnotesize{(Colour online).  The left plot shows FES estimates for $-2 \Delta$ obtained using all bond dimensions from $D=32$ to $D=64$. The red line denotes the estimates themselves, i.e. the slopes obtained by linear interpolation of $\log(G(s \mu_2(D)))$ vs. $\log(  \mu_2(D))$, for different scales $s$. The black lines denote the errors in the estimates, given by the confidence intervals for the slopes at $99.73\%$ confidence level. The right plot displays the combination of the Direct Approach and FES. The final estimate for the exponent (its position is approximately indicated by the red circle), is given by the most accurate FES estimate in range of scales bounded on one side by the largest value of $s$ at which the two approaches agree below the correlation length, i.e. for $s<1$, and on the other by  $s=\infty$. The green line in both plots denotes the exact $-2 \Delta$ value obtained using the Bethe Ansatz.}}
\label{fig:LL_field_scaling_direct_combined}
\end{figure*}

As discussed in Section \ref{sec:D_scaling},  the Direct Approach on its own is not useful for obtaining good estimates for the exponents. As can be seen in Figure \ref{fig:LL_psi_disconn_correlator}, power law decay  for the cMPS approximation to the field-field two-point correlation function at some fixed bond dimension $D$ is captured well beyond some short distance at which non-universal effects are present, up to approximately the correlation length,  beyond which exponential decay takes over.  The left plot in Figure \ref{fig:LL_field_direct_exponent_combined} explicitly demonstrates that estimates for $-2\Delta$, computed from the derivative of $\log(G(x))$ vs. $\log (x)$, are completely off at distances shorter than some cutoff, and also beyond the correlation length. The problem with the Direct Approach therefore lies both in the difficulty of determining the window in which estimates are reliable, and in the lack of any method to determine the error in the estimates.

One could attempt to work around these obstacles by obtaining estimates using a range of bond dimensions, and scanning for the scale  at which their spread is minimal.  For the case at hand, using all bond dimensions $D$ between $32$ and $64$, we obtain the result shown in the right plot of Figure \ref{fig:LL_field_direct_exponent_combined}.  In this case the true value is actually captured by this method, but this turns out to be a lucky accident. The approach fails for most operators we considered in the paper. 

The overlay of the Direct Approach, using all bond dimensions $D$ between $32$ and $64$,  and FES   is displayed in the right plot of Figure \ref{fig:LL_field_scaling_direct_combined} and demonstrates how our best estimate, $2\Delta = 0.1667^{+0.0005}_{-0.0005} $, at $99.73\%$ confidence level, is obtained (see Table \ref{tab:LL_results}). The region of overlap between the two below $s=1$ gives an upper bound for $s_0$, i.e. the scale we are able to scan down to without encountering short distance/cutoff effects. The best estimate is then determined to be at $s=0.66$; the interpolation at $s=0.66$  is depicted in Figure \ref{fig:LL_field_logcorr_vs_logpos_with_fit}.

It is instructive to think of the plots in Figures  \ref{fig:LL_psi_disconn_correlator}  and \ref{fig:LL_field_logcorr_vs_logpos_with_fit} in terms of appropriate intersections of the two-dimensional surface displayed in Figure \ref{fig:LL_field_plot3d}.
   \begin{figure*}[!htb]
\centering
\includegraphics[width=16cm]{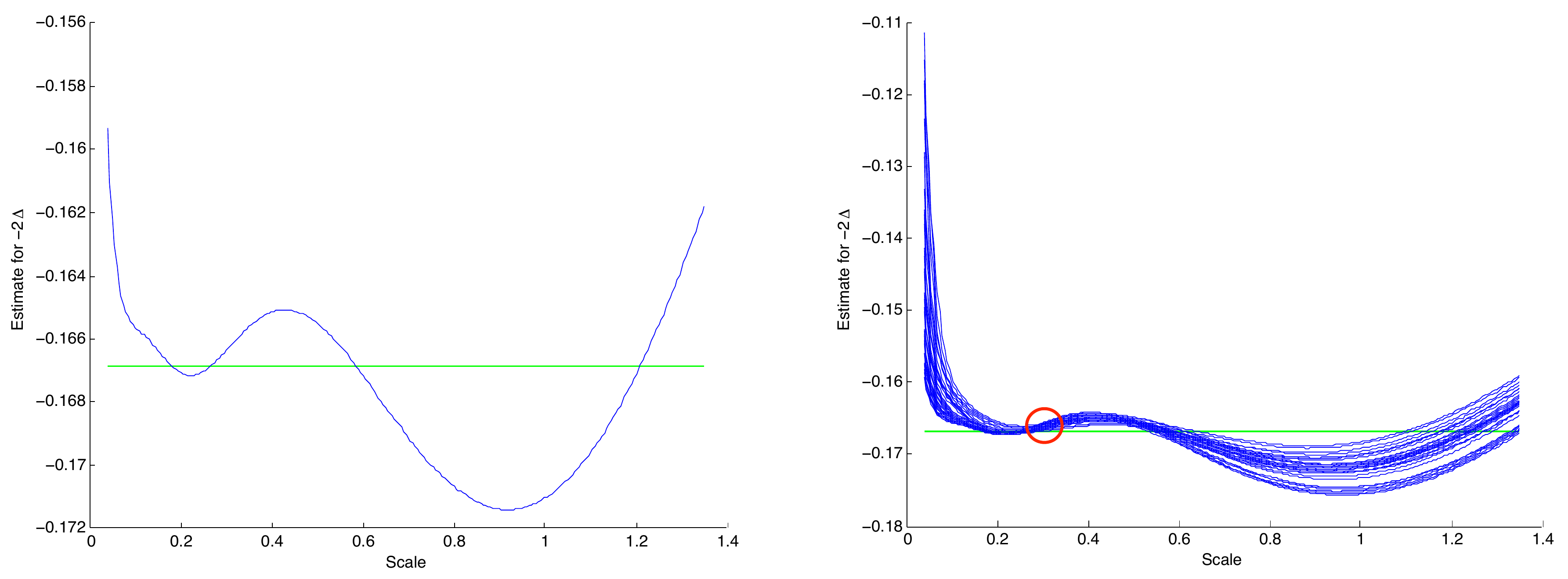}
\caption{\footnotesize{(Colour online.) The left plot shows estimates for $-2\Delta$ obtained directly from the cMPS approximation to the field-field Lieb-Liniger correlator computed at the maximum available bond dimension $D=64$. The $x$-axis variable is the scale $s$, $x = s \mu_2(D)$, and the estimate for $-2\Delta$ is obtained by computing the derivative of $\log(G(x))$ vs. $\log (x)$.   The right plot combines data for all bond dimensions $D$ ranging from $32$ to $64$. In both plots the green line denotes the exact value as obtained using the Bethe Ansatz.  The right plot demonstrates that the region where the spread of values is minimal, denoted by the red circle, actually captures the true value. In general this approach does not provide a reliable method for estimating critical exponents, contrary to what the current example indicates.} }\label{fig:LL_field_direct_exponent_combined}
\end{figure*}

  \newpage

\section{Central charge estimates from $D$-scaling }
\label{app:kappa}
 
 \begin{table*}[!htb]
\centering
\begin{tabular}{| c |c |c |c | c |  c| c| }
\hline
Model &   Slope &                Slope &                           Predicted& $c$   Estimate  & $c$  Estimate    \\  
  &      $99.73\%$ conf.    &  $95 \%$ conf.    &   Slope   &     $99.73\%$ conf.   &   $95\%$ conf.  \\ 
\hline \hline 
Lieb-Liniger &  $1.30^{+0.03}_{-0.03}$ & $1.295^{+0.020}_{-0.019}$ & $\kappa = 1.3441... $ &  $1.06^{+0.04}_{-0.04}$ &  $1.061^{+0.027}_{-0.025}$ \\  \hline
Relativ. Boson &   $1.26^{+0.04}_{-0.04}$ & $1.256^{+0.023}_{-0.023}$ & $\kappa = 1.3441... $ &  $1.12^{+0.06}_{-0.05}$ &  $1.12^{+0.03}_{-0.03}$ \\  \hline
Quantum Ising &   $1.91^{+0.05}_{-0.05}$ & $1.91^{+0.03}_{-0.03}$ & $\kappa = 2.0343... $ &  $0.558^{+0.027}_{-0.026}$ &  $0.558^{+0.016}_{-0.017}$ \\ 
\hline
\end{tabular}
\caption{\footnotesize{ Summary of estimates for $\kappa  $ obtained   from the scaling of $\log (\mu_2(D))$ vs. $\log(D)$  using all $D$ in the range  $32  \leq D \leq 64$  for the Lieb-Liniger, massless relativistic boson, and critical quantum Ising models.  The analytic relation $\kappa = 6/\left( c \left( \sqrt{\frac{12}{c}} + 1 \right) \right) $ is not reproduced very well, and the related central charge estimates, given in the two rightmost columns, are therefore also inaccurate. }}
\label{tab:kappa_results}
\end{table*}
 
This Appendix presents  estimates for the central charge $c$ for the  Lieb-Liniger, massless relativistic boson, and critical quantum Ising model, obtained by scaling directly with respect to the bond dimension $D$.  The exact central charge for the two field theories is $c=1$, and for the quantum Ising model $c=1/2$.  The exponent $\kappa$ is determined from  scaling $\log (\mu_2(D))$ vs. $\log(D)$, $\mu_2(D)\sim D^{\kappa}$ (see the discussion around Eq. (\ref{eq:kappa_def}) in Section \ref{sec:scalinghypothesis}). A set of estimates for $c$ is then obtained using the analytic relation $\kappa = 6/\left( c \left( \sqrt{\frac{12}{c}} + 1 \right) \right)$ (Table \ref{tab:kappa_results}). Further estimates are obtained from scaling the entropy $S$ with $\log (D)$ (Tables \ref{tab:c_half_line_results_kappa} and \ref{tab:c_LL_results_interval_kappa}), both after making use of the $\kappa(c)$ relation, and also while keeping $\kappa$ as a free parameter (i.e. using the values obtained in Table \ref{tab:kappa_results}).    The $\kappa(c)$ relation is expected to be only approximately true, and the inaccuracy of the results based on this relation demonstrates that it does not hold very accurately in the region of bond dimensions  $32  \leq D \leq 64$ used for the scalings. The accuracy of the results obtained with $\kappa$ as a free parameter is much better, but the error bars are larger than those obtained when scaling w.r.t. $\mu_2(D)$, as in Section \ref{section:examples}
 of this paper.  Results in Tables \ref{tab:c_half_line_results_kappa}  and \ref{tab:c_LL_results_interval_kappa} should be compared with results obtained by scaling $S$  directly with $\mu_2(D)$, as presented in Tables \ref{tab:c_half_line_results} and  \ref{tab:c_LL_results_interval}.
\begin{table*}[!htb]
\centering
\begin{tabular}{|c |c |c |c | c |  c| c| }
\hline
Model &   Slope &                 Predicted& $c$   Estimate  & $c$  Estimate with   \\  
   &    $99.73\%$ conf.       &   Slope   &     using $\kappa(c)$   &   $\kappa$ a Free Parameter  \\ 
\hline \hline 
Lieb-Liniger & $0.212^{+0.008}_{-0.008}$  & $\frac{1}{\left( \sqrt{\frac{12}{c}} +1 \right)} = 0.22401...$ &$0.87^{+0.09}_{-0.08}$ & $0.98^{+0.04}_{-0.04}$ \\ \hline
Relativ. Boson & $0.215^{+0.005}_{-0.005}$  & $\frac{1}{\left( \sqrt{\frac{12}{c}} +1 \right)} = 0.22401...$ &$0.90^{+0.05}_{-0.05}$ & $1.024^{+0.024}_{-0.024}$ \\ \hline
Quantum Ising &  $0.158^{+0.006}_{-0.006}$  & $\frac{1}{\left( \sqrt{\frac{12}{c}} +1 \right)}  = 0.169521...$ &$0.42^{+0.04}_{-0.04}$ & $0.496^{+0.019}_{-0.018}$ \\ \hline
\end{tabular}
\caption{\footnotesize{ Summary of central charge estimates for the Lieb-Liniger, massless relativistic boson, and critical quantum Ising models obtained from the scaling of the entropy $S$ vs. $\log(D)$. We present results obtained both using the conjectured dependance of $\kappa$ on $c$ (\ref{eq:kappa_def}), and also when keeping $\kappa$ a free parameter, that is, using the values obtained in Table \ref{tab:kappa_results}.  }}
\label{tab:c_half_line_results_kappa}
\end{table*}
\begin{table*}[!htb]
\centering
\begin{tabular}{|c |c |c |c | c |  c| c| }
\hline
Model &    Slope &                Predicted& $c$   Estimate  & $c$  Estimate with    \\  
    &   at $99.73\%$ conf.       &   Slope   &    using $\kappa(c)$  &  $\kappa$ a Free Parametner  \\ 
\hline \hline 
Lieb-Liniger & $0.423^{+0.012}_{-0.011}$  &  $\frac{2}{\left( \sqrt{\frac{12}{c}} +1 \right)} = 0.448018...$ &$0.90^{+0.06}_{-0.06}$ & $0.976^{+0.025}_{-0.028}$ \\ \hline
Relativ. Boson &  $0.423^{+0.010}_{-0.010}$  & $\frac{2}{\left( \sqrt{\frac{12}{c}} +1 \right)} = 0.448018...$ &$0.86^{+0.06}_{-0.05}$ & $1.007^{+0.024}_{-0.025}$ \\ \hline
\end{tabular}
\caption{\footnotesize{ Summary of central charge estimates for the Lieb-Liniger and massless relativistic boson models obtained by scaling the entanglement entropy $S$ of an interval at scale $s=0.1$ vs. $\log(D)$. We again have two sets of results, one obtained while making use of the conjectured dependance of $\kappa$ on $c$ (\ref{eq:kappa_def}), and the other by keeping $\kappa$ a free parameter.  The linearity of the fits is improved compared to those displayed in Table \ref{tab:c_half_line_results_kappa}}, however the estimates based on $\kappa(c)$ still miss the true values.   }
\label{tab:c_LL_results_interval_kappa}
\end{table*}

\end{small}

\newpage

\bibliography{bibliography}

\end{document}